\documentclass[aps,pra,twocolumn,showpacs,superscriptaddress,preprintnumbers,amsmath,amssymb,floatfix]{revtex4-1}

\usepackage{graphicx}
\usepackage{diagbox}
\usepackage{subfigure}
\usepackage{dcolumn}
\usepackage{bm}
\usepackage{color}
\usepackage{ulem}
\usepackage{array}
\usepackage{stfloats}
\newcolumntype{P}[1]{>{\centering\arraybackslash}p{#1}}

\newcommand{\etal}{{\it et al.\! }}
\newcommand{\pr}[4]{Phys. Rev. #1 {\bf #2}, #3 (#4)}
\newcommand{\hedp}[3]{High Energy Density Phys. {\bf #1}, #2 (#3)}
\newcommand{\astropj}[3]{Astrophys. J. {\bf #1}, #2 (#3)}

\newcommand{\physfluid}[3]{Phys. Fluids {\bf #1}, #2 (#3)}

\newcommand{\rmboth}[3]{{#1}^{{\rm #2}}_{{\rm #3}}}
\newcommand{\rmnumb}[2]{{#1}_{\rm #2}}
\newcommand{\rmupper}[3]{{#1}^{{\,\rm #2}}_{#3}}
\newcommand{\ket}[1]{| #1 \rangle}

\newcommand{\mean}[2]{\langle\, #1 \, \rangle_{\rm #2}}

\newcommand{\bks}[1]{\left( #1 \right)}
\newcommand{\curlybra}[1]{\left\{ #1 \right\}}
\newcommand{\squarebra}[1]{\left[ #1 \right]}

\newcommand{\biggcurlybra}[1]{\bigg\{ #1 \bigg\}}
\newcommand{\Bigcurlybra}[1]{\Big\{ #1 \Big\}}

\newcommand{\impart}[0]{{\rm Im}\, }
\newcommand{\repart}[0]{{\rm Re}\, }

\newcommand{\kbt}[1]{\rmnumb{k}{B} T_{#1}}

\newcommand{\expf}[0]{\mathrm{exp}\!}

\begin{document}


\title{Ionization potential depression and ionization balance in dense plasmas}

\author{Chengliang Lin}
\email{cllin@gscaep.ac.cn; chengliang1064@gmail.com}
\affiliation{Graduate School of China Academy of Engineering Physics, Beijing 100193, P. R. China}
\affiliation{Institut f\"ur Physik, Universit\"at Rostock, 18051 Rostock, Germany}

\date{\today}

\begin{abstract}
Theoretical modelling of ionization potential depression and the related ionization equilibrium in dense plasmas, 
in particular in warm/hot dense matter,
represents a significant challenge due to ionic coupling and electronic degeneracy effects. 
We present a quantum statistical model based on dynamical structure factors
for the ionization potential depression, where quantum exchange and dynamical correlation effects in plasma
environments are consistently and systematically taken into account in terms of the concept of self-energy. 
Under the condition of local thermodynamic equilibrium, the charge state distribution (or ionic fraction) 
characterized by the ionization balance is obtained by solving the coupled Saha equations. Calculations for 
the ionization potential depression of different chemical elements are performed with the electronic
and ionic structure factors. The ionic structure factors are determined by solving the Ornstein-Zernike equation in combination with the hypernetted-chain equation. As a further application of our approach, 
we present results for the charge state distribution of aluminium plasmas at several temperatures and densities.

\end{abstract}

%

\maketitle

\section{Introduction}
\label{introduction}

Within the plasma community, one of the well-known many-body effects is ionization potential depression (IPD) 
or continuum lowering. The IPD significantly alters the ionization balance and the corresponding charge state distribution, 
which has a strong influence on transport, optical, and thermodynamic properties of the system. This is of 
essential importance not only for plasma physics but also for astrophysics, planetary science, and solid 
state physics. However, an accurate description of IPD in an interactive many-body
environment is extremely complicated, because strong correlations and quantum degeneracy have to be taken into
account consistently. Different semi-empirical models have been proposed for the IPD. In particular, 
the Ecker-Kr\"oll (EK)~\cite{EK63} and 
the Stewart-Pyatt (SP) model~\cite{SP66} have been widely applied in plasma physics for simulations of physical 
properties and the analysis of experimental observations. 

With newly developed experimental facilities, it has 
become possible to explore warm/hot dense matter and materials in the high-energy density regime, whereby 
the physical systems are found to be strongly coupled and nearly degenerate. The experimental outcomes can 
be used to benchmark physical assumptions and theoretical models commonly applied within plasma physics. 
Over the last few years, new experiments related to the IPD have been performed with high-intense laser 
beams in LCLS at SLAC National Accelerator Laboratory~\cite{ciricosta12,ciricosta16}, 
at ORION in the UK~\cite{Hoarty13}, and in NIF at Lawrence Livermore
National Laboratory~\cite{Fletcher14,Kraus16}. These precise experiments have revealed the lack of a consistent
picture about IPD, since no self-consistent explanation for these experiments can be drawn from any of the commonly accepted IPD models.
Therefore, a more fundamental understanding of the phenomenon IPD in warm/hot dense regime is required.
Recently, attempts for a better understanding of the new experimental data have been made using 
different numerical approaches and simulation
methods. Two-step Hartree-Fock calculations~\cite{STJZS14}, simulations based on the finite-temperature 
density functional theory~\cite{VCW14,Hu17}, classical
molecular dynamics simulations~\cite{Calisti15}, and Monte-Carlo simulations ~\cite{Stransky16} have been worked out.
Although these state-of-the-art modelling methods are very successful to explain properties of warm dense matter, they
are restricted due to the large computational cost. Besides these simulations methods, other analytic
improvements have also been proposed, such as fluctuation model~\cite{IS13} and atomic-solid-plasma model~\cite{Rosmej18}.

In view of the demands from the theoretical as well as experimental aspect, we have, starting from a
quantum statistical theory, derived an analytic self-consistent approach to IPD, which is demonstrated to
be valid in a wide range of temperatures and densities~\cite{LRKR17}.  The influence of the surrounding plasma on an embedded ion/atom is described by the self energy (SE) which contains the dielectric 
function~\cite{KKER86,KSKB05}. The dielectric function can be expressed in terms of the dynamic
structure factor (SF) according to the fluctuation-dissipation theorem. Since the dynamic SF comprises
the time and space correlations of particles in the plasma, a microscopic understanding of the IPD in
this quantum statistical approach is obvious. The developed quantum statistical model for IPD is 
based on the assumption of a two-component plasma model with ionic and electronic
subsystem. In a realistic physical problem, the investigated system usually consists of different ion
species~\cite{WCK17}, for example, H-C mixture in laboratory physics~\cite{Kraus16,LR12}. 
Because of the large mass and charge
asymmetry, the light element moves more liberally between the highly charged,
strongly correlated heavy components, which generally causes an additional dynamical screening effect 
for the calculation of IPD.

In this work we give a detailed description of our quantum statistical approach for IPD,
which is based on the model developed in the previous work~\cite{LRKR17} and is extended to describe 
a multicomponent plasma. The present work is organized as follows: in Sec.~\ref{SE&IPD}, we outline 
the basis of single-particle SE in the GW approximation, where G denotes the dressed Green's function and
W indicates the dynamically screened interaction potential. In terms of the SE, definition of IPD within the quantum statistical theory is introduced in Sec.~\ref{DefineIPD}. In the subsequent Sec.~\ref{IPDandSF}, we demonstrate that the IPD is 
connected to charge-charge dynamical SF, which describes dynamical correlation effects in plasmas.
Quantum exchange (statistical) correlations including the Fock contribution and the Pauli blocking 
effect are also taken into account in this section. Using the effective ionization potential defined within the
developed IPD model, the derivation of the coupled Saha equations in the chemical picture is discussed in 
Sec.~\ref{CPandSE}.
In the high-temperature ideal plasma limit, the Debye-H\"uckel model 
is exactly reproducible from our approach, as shown in Sec.~\ref{DHlimit}. As applications of our method, 
the IPD for different chemical elements are investigated in Sec.~\ref{IPDcalculation}, where comparisons
with experimental observations~\cite{ciricosta16} are performed. Then we apply the developed IPD model to calculate the charge 
state distribution of Al plasmas corresponding to the thermodynamical conditions in experiments of 
Hoarty \etal~\cite{Hoarty13} in Sec.~\ref{ionicFraction}. Finally, conclusions are drawn in Sec.~\ref{Conclusion}.

\section{Self energy and ionization potential depression}
\label{SE&IPD}

\subsection{Plasma parameters}
We consider a multicomponent mixture consisting of $\rmnumb{N}{e}$ free electrons with charge $-\, e$ and 
mass $\rmnumb{m}{e}$ as well as $N_\gamma$ ions of different species $\gamma$ with charge $z_\gamma\, e$ and 
mass $m_\gamma$ in a volume $V$. $e$ is the elementary charge. The partial particle number density 
for ion species $\gamma$ is $n_\gamma = N_\gamma/V$ and the total particle number density for all ions is 
$\rmnumb{n}{heavy} = \sum_\gamma n_\gamma$. The corresponding number concentration is then given by 
$x_\gamma = n_\gamma / \rmnumb{n}{heavy}$. According to the charge neutrality, the electron density is 
$\rmnumb{n}{e} = \sum_\gamma z_\gamma\, n_\gamma = \bar z \, \rmnumb{n}{heavy}$ with the mean ionization
degree of ions $\bar z = \sum_\gamma z_\gamma\, x_\gamma$. Additionally, we introduce the effective
charge number of plasma ions 
$\rmnumb{z}{p} = \sum_\gamma z^2_\gamma\, x_\gamma / \bks{\sum_\gamma z_\gamma\, x_\gamma} = \mean{z^2}{}/\bar z $,
which effectively describes the plasma as a whole and therefore regards the ionic perturbers in plasma as a single ionic species.

In a many-body environment the motion of particles is correlated with the motion of their nearby particles. 
The coupling strength of such correlation is represented by the dimensionless plasma parameter
\begin{equation}
\Gamma_{cd} 
= \frac{z_c z_d \, e^2 }{4\pi \varepsilon_0\, a_{cd}\, \kbt{cd}},
\end{equation}
which is taken as the ratio of the average unscreened interaction potential between type $c$ and type $d$
to the thermodynamic kinetic energy characterized by the temperature $T_{cd}$. 
In general, ions and electrons in multicomponent charged particle systems can have 
different temperatures with $T_{\mathrm{ee}}$ and $T_{\mathrm{ii}}$, respectively.
The electron-ion interaction temperature $T_{\mathrm{ei}}$  has been constructed with different ansatzes, see Refs.~\cite{GRHGR07,HFBKRY17}. The averaged interparticle distance reads
\begin{equation}
 a_{cd} = \squarebra{ \frac{4\pi \bks{ n_c + n_d }/2}{3} }^{-1/3}.
\end{equation}
Additionally, the quantum degeneracy for the electron subsystem is defined via the ratio 
of the thermodynamic kinetic energy $\kbt{\mathrm{e}}$ and the Fermi energy 
$\rmnumb{E}{F} = \hbar^2 \bks{3\pi^2 \rmnumb{n}{e}}^{2/3}/\bks{2\rmnumb{m}{e}}$ as follows
\begin{equation}
 \rmnumb{\theta}{ee} = \frac{\kbt{\mathrm{ee}}}{\rmnumb{E}{F}}.
\end{equation}

\subsection{Single-particle self energy in GW approximation}
We calculate the single-particle SE (SPSE) in the GW-approximation. In the Lehman 
representation~\cite{KKER86}, the Green's function for species $c$ in the energy-momentum space, 
i.e. $G_c(\mathbf{p},z_\mu)$, can be expressed in terms of the spectral function $A_c(\mathbf{p},\omega)$
\begin{equation}
 G_c(\mathbf{p},z_\nu) = \int_{-\infty}^\infty \frac{d \omega}{2\pi}\, \frac{A_c(\mathbf{p},\omega)}{z_\nu - \omega},
\end{equation}
where $z_\nu = \pi\nu/\bks{\hbar \beta}$ is the Matsubara frequencies with 
$\nu = \pm 1, \pm 3, \cdots$ for fermions and $\nu = 0,\pm 2, \pm 4, \cdots$ for bosons.
Similarly, the screened interaction potential $W(\mathbf{k},\omega)$ can be written in the spectral 
representation via the inverse dielectric function
\begin{align}
 W_{ab}(\mathbf{k},\omega) = V_{ab}(\mathbf{k}) \, \squarebra{ 1 +  \int_{-\infty}^\infty \frac{d \omega_1}{\pi}\,
 \frac{ \impart \varepsilon^{-1}(\mathbf{k},\omega_1) }{\omega - \omega_1}}
\end{align}
with the Coulomb interaction $ V_{ab}(\mathbf{k}) = z_a z_b e^2 / \bks{\varepsilon_0 \, k^2}$.
The inverse dielectric function describes the response of a many-body system to an external perturbation.
Then the SPSE in GW approximation 
\begin{equation}\label{singleSE}
 \Sigma_c(\mathbf{p},z_\nu)  = -\frac{1}{\beta} \sum_{\mathbf{k},\omega_\mu}
 G_c(\mathbf{p-k},z_\nu-\omega_\mu)\,  W_{cc}(\mathbf{k},\omega_\mu) 
\end{equation}
can be decomposed into a Hartree-Fock (HF) contribution due to quantum exchange effects and a
correlation one because of dynamical interactions
\begin{align}\label{intContrib}
 \Sigma_c(\mathbf{p},z_\nu) =  \rmupper{\Sigma}{HF}{c}(\mathbf{p},z_\nu) 
 + \rmupper{\Sigma}{corr}{c}(\mathbf{p},z_\nu),
\end{align}
where the HF SE is given by
\begin{align}\label{HFcontrib}
\rmupper{\Sigma}{HF}{c}(\mathbf{p}) & = -\frac{1}{\beta} \sum_{\mathbf{k},\omega_\mu}\,
 V_{cc}(\mathbf{k})\,
 \int_{-\infty}^\infty \frac{d \omega}{2\pi}\, \frac{A_c(\mathbf{p-k},\omega)}{z_\nu -\omega_\mu - \omega} ,
\end{align}
and the correlation part of the SPSE reads
\begin{align}\label{corrcontrib}
 \rmupper{\Sigma}{corr}{c}(\mathbf{p},z_\nu) & =
 -\frac{1}{\beta} \sum_{\mathbf{k},\omega_\mu}\,V_{cc}(\mathbf{k})\,
 \int_{-\infty}^\infty \frac{d \omega}{2\pi}\, \frac{A_c(\mathbf{p-k},\omega)}{z_\nu -\omega_\mu - \omega} \nonumber \\
 & \qquad \times \int_{-\infty}^\infty \frac{d \omega_1}{\pi}\,
 \frac{ \impart \varepsilon^{-1}(\mathbf{k},\omega_1) }{\omega_\mu - \omega_1} .
\end{align}
In the following, the sum over momentum $\mathbf{k}$, i.e. $\sum_{\mathbf{k}}$, is replaced by the integral $\int d^3\mathbf{k}/(2\pi)^3$. Performing the summation over the Matsubara frequencies $\omega_\mu$ yields 
\begin{align}
 &  \rmupper{\Sigma}{corr}{c}(\mathbf{p},z_\nu \!)  \! = \!\!\!\int\!\!\! \frac{d^3 \mathbf{k}}{(2\pi)^3} 
 \! V_{cc}(\mathbf{k})\!\!\!
 \int_{-\infty}^\infty \!\!\!\frac{d \omega_1}{2\pi}\! A_c(\mathbf{p\!-\!k},\omega_1) M(\mathbf{k},z_\nu,\omega_1) \label{Intint} \\
&  \rmupper{\Sigma}{HF}{c}\!(\mathbf{p})  \! = \!\!\int\!\! \frac{d^3 \mathbf{k}}{(2\pi)^3} V_{cc}(\mathbf{k})\!\!
 \int_{-\infty}^\infty \!\frac{d \omega}{2\pi}\! A_c(\mathbf{p-k},\omega)
  \Bigcurlybra{\!1\! -\!n_c(\omega)\! }, \label{HFint} 
\end{align}
where the distribution function $n_c(\omega)$ reads
\begin{equation}\label{distributionFunc}
n_c(\omega) = \frac{1}{\expf\squarebra{\beta\bks{\hbar\omega - \mu_c}}\pm 1}
\end{equation}
with the upper sign $+$ for fermions denoted as $\rmnumb{n}{F}(\omega)$, and the lower sign $-$ for bosons with
$\rmnumb{n}{B}(\omega)$. $ \mu_c$ is the chemical potential of species $c$. The dynamical effects within mediums are described by the function $ M(\mathbf{k},z_\nu,\omega_1)  $ that is given by
\begin{align}
 M(\mathbf{k},z_\nu,\omega_1)  = & \int_{-\infty}^{\infty}  \frac{d\omega_2}{\pi} 
\,\impart \varepsilon^{-1}(\mathbf{k},\omega_2) \\ &  \times
 \frac{ \rmnumb{n}{B}(\omega_2) + 1 - \rmnumb{n}{F}(\omega_1) }{z_\nu - \omega_1 - \omega_2 } \nonumber .
\end{align}
The HF SE, Eq.~\eqref{HFint}, has no dependence on the frequency $z_\nu$ and is a real quantity.
The first contribution of the HF SE is denoted as Hartree term which vanishes for a homogeneous system
because of charge neutrality~\cite{KKER86}. The second contribution is the so-called Fock term, which arises from
exchange correlation of identical particles and has no classical counterpart. We only calculate
the Fock term in this work, but still refer this contribution as Hartree-Fock 
\begin{equation}
 \rmupper{\Sigma}{HF}{c}(\mathbf{p})\! =\! -\sum_{\mathbf{k}}V_{cc}(\mathbf{k})\!
 \int_{-\infty}^\infty \frac{d \omega}{2\pi} \,n_c(\omega) A_c(\mathbf{p-k},\omega).
\end{equation}

The correlation contribution, i.e. Eq.~\eqref{Intint}, can be split into a real
and an imaginary part by means of the analytic continuation (i.e. Wick rotation) 
$z_\nu \rightarrow z = \omega + i \delta$ with $\delta\rightarrow 0^+$~\cite{KKER86,KSKB05,SL13}.
After the analytic continuation the function $M(\mathbf{q},\omega,\omega_1)$ can be rewritten as 
\begin{align}\label{Mfunction}
& M(\mathbf{q}, \omega ,\omega_1)  \\
= & {\cal P} \int_{-\infty}^{\infty}  \frac{d\omega_2}{\pi} 
\,\impart \varepsilon^{-1}(\mathbf{q},\omega_2) \cdot 
 \frac{ \rmnumb{n}{B}(\omega_2) + 1 - \rmnumb{n}{F}(\omega_1) }{\omega  - \omega_1 - \omega_2 } \nonumber \\
&  - i\, \impart \varepsilon^{-1}(\mathbf{q},\omega-\omega_1)\cdot \curlybra{
  \rmnumb{n}{B}(\omega-\omega_1) + 1 - \rmnumb{n}{F}(\omega_1) } \nonumber 
\end{align}
with the use of Dirac's identity 
$\frac{1}{x\, \pm\, i\, 0^+} = {\cal P} (\frac{1}{x}) \mp i\pi \delta(x) $ (${\cal P}$ denotes the principal value).
Generally, we have $\rmnumb{n}{F}(\omega_1)<<1$ for the non-degenerate ions, in particular, for the ion involved in
the ionization process. Obviously, the correlation contribution of the SPSE, i.e. Eq.~\eqref{Intint}, can be decomposed into a
real part (related to the shift of eigenstates) and an imaginary part (connected to the broadening of eigenstates).

\subsection{Definition of ionization potential depression within the quantum statistical theory}
\label{DefineIPD}

Generally, the IPD in a medium is defined as the change of 
the ionization potential with respect to the isolated case. It can be extracted from the effective
binding energy, which is obtained by solving the Schr{\"o}dinger equation with an effective interaction 
potential. The IPD acquired from the solution of the standard Schr{\"o}dinger equation (i.e.\! eigenenergies 
for both scattering and bound eigenstates) is a real quantity. However, the quantum eigenstates of the 
investigated system in a many-body environment are visibly broadened due to the fast oscillating part of 
the microfield generated by the surrounding charged particles. In order to account for both the shift and 
the broadening of quantum eigenstates, the Bethe-Salpeter equation (or the in-medium Schr{\"o}dinger equation) 
has to be solved~\cite{KKER86}. Therefore, the commonly defined IPD can be extended to a complex quantity,
which includes the standard IPD  and an additional contribution due to
broadening effects.

Alternatively, the problem of generalized IPD (GIPD)
can also be tackled within the Green's function technique for the quantum statistical theory~\cite{LRKR17}.
In the framework of quantum statistical theory, the modifications of the atomic/ionic properties 
are described by the SE. In the chemical picture, the investigated system undergoing 
ionization is treated as different ionic species before (ion $\alpha$) and after the ionization
(ion $\alpha+1$ plus an ionized electron). 
Apparently, the SE of the investigated system before and after the ionization is 
significantly changed, while the surrounding environment is assumed to be not changed during 
the ionization process if the relaxation effect is negligible. Under this assumption, the GIPD 
$\rmboth{\cal I}{\text{\tiny GIPD}}{\alpha} $ can be defined via the difference between the 
SEs of the corresponding investigated system before and after the ionization.
In the following, the letters $(a,b,\cdots)$ in the subscripts and in the summation 
represent both ions and electrons. If only ionic species are involved in equations, the
Greek letters $(\alpha,\beta,\cdots)$ will be used.

The SPSE of species $c$ is expressed via the dressed propagator (Green's function) 
$G_c(\mathbf{p},\omega)$ and the screened interaction potential $W_{cc}(\mathbf{k},\omega)$~\cite{KKER86,KSKB05} (see also Eq.~\eqref{singleSE})
\begin{equation}
 \Sigma_c(\mathbf{p},\omega)  = -\frac{1}{\beta} \sum_{\mathbf{k},\omega_1}
 G_c(\mathbf{p-k},\omega-\omega_1)\cdot  W_{cc}(\mathbf{k},\omega_1).
\end{equation} 
Then the frequency- and momentum-dependent GIPD is given by
\begin{equation}\label{IPDdefine}
\rmboth{\cal I}{\text{\tiny GIPD}}{\alpha}\!(\mathbf{p},\omega)\!\! =\! \!\Sigma_{\alpha}(\mathbf{p},\omega) 
 \!- \!\Bigcurlybra{\! \Sigma_{\alpha+1}\!(\mathbf{p},\omega)+\! \rmboth{\Sigma}{e}{ionized}\!(\mathbf{p},\omega)\! }.
\end{equation}
The commonly defined IPD is described by the real part of the above introduced GIPD
\begin{align}\label{IPDreal}
{\cal I}_\alpha(\mathbf{p},\omega)  = \repart \, \rmboth{\cal I}{\text{\tiny GIPD}}{\alpha}(\mathbf{p},\omega) ,
\end{align}
and the broadening of the IPD is given as
\begin{align}\label{IPDimag}
{\cal B}_\alpha(\mathbf{p},\omega)  = \impart \, \rmboth{\cal I}{\text{\tiny GIPD}}{\alpha}(\mathbf{p},\omega).
\end{align}
The electronic SE $ \rmboth{\Sigma}{e}{ionized}(\mathbf{p},\omega)$ contains two contribution:
the SPSE of the ionized electron as a free particle $(\rmnumb{\Sigma}{e})$ and 
energy modification of the ionized electron in its parent ion as well as
the restriction of phase space occupation in the bound-free ionization process, i.e. 
\begin{equation}\label{ionEEalpha}
 \rmboth{\Sigma}{e}{ionized} = \rmnumb{\Sigma}{e} + \rmboth{\Sigma}{e}{bf}.
\end{equation}
Concerning bound-free transition of the ionized electron, the following many-particle effects have to be
considered in the SE $\rmboth{\Sigma}{e}{bf}$:(1) energy levels of the bound states are shifted due 
to dynamical interaction (collisional shift) and quantum exchange effect (Fock shift) between bound electron 
and free electrons in plasmas; 
(2) bound electrons can not be ionized to those states that are already occupied by the free electrons 
(known as Pauli blocking);
(3) the energy levels are also broadened because of random collision of bound electron with its surrounding
charged particles (known as pressure broadening).  
In this work, we only consider the Pauli blocking and the Fock shift of bound states in 
$\rmboth{\Sigma}{e}{bf}$, i.e.
\begin{equation}
 \rmboth{\Sigma}{e}{bf} \approx \rmboth{\Delta}{PF}{bf}
 =  \rmboth{\Delta}{Pauli}{bf} +  \rmboth{\Delta}{Fock}{bf}.
\end{equation}
Then the GIPD can be defined as
\begin{equation}\label{defIPD}
 \rmboth{\cal I}{\text{\tiny GIPD}}{\alpha} = \Sigma_{\alpha}
- \bks{ \Sigma_{\alpha+1}+ \rmboth{\Sigma}{}{e} +\rmboth{\Delta}{PF}{bf} },
\end{equation}
where the momentum- and frequency-dependence are suppressed. 
The basic quantity describing the GIPD $\rmboth{\cal I}{\text{\tiny GIPD}}{\alpha}$ 
is the SPSE $\Sigma_c$, which can be evaluated within different approximations. In the next section, 
we will discuss the G$_0$W approximation for the SPSE $\Sigma_c$, the Pauli blocking of the ionized electron
$\rmboth{\Delta}{Pauli}{bf}$, and the Fock shift of bound states $\rmboth{\Delta}{Fock}{bf}$.

Based on the SPSE, we can define the frequency- and momentum-dependent GIPD in this section and its 
reduced version in the next section. Inserting Eq.~\eqref{intContrib} in combination with 
Eq.~\eqref{HFcontrib} and Eq.~\eqref{corrcontrib} into Eq.~\eqref{defIPD} yields
\begin{align}\label{corrcontrib0}
 \rmboth{\cal I}{\text{\tiny GIPD}}{\alpha}(\mathbf{p},\omega) = \rmupper{\cal I}{HF}{\alpha}(\mathbf{p}) 
+\rmupper{\cal I}{corr}{\alpha}(\mathbf{p},\omega) -\rmboth{\Delta}{PF}{bf}(\mathbf{p})
\end{align}
with the Hartree-Fock contribution
\begin{align}
\rmupper{\cal I}{HF}{\alpha}(\mathbf{p})  = \rmupper{\Sigma}{HF}{\alpha}(\mathbf{p})
- \rmupper{\Sigma}{HF}{\alpha+1}(\mathbf{p}) - \rmupper{\Sigma}{HF}{\mathrm{e}}(\mathbf{p}) ,
\end{align}
and correlation contribution from dynamical interactions
\begin{align}\label{intContrib22}
\rmupper{\cal I}{corr}{\alpha}\! (\mathbf{p},\omega) \! =\!
 \rmupper{\Sigma}{corr}{\alpha}\!(\mathbf{p},\omega)\!
 - \! \rmupper{\Sigma}{corr}{\alpha+1}\!(\mathbf{p},\omega)\!
 - \! \rmupper{\Sigma}{corr}{\mathrm{e}}\!(\mathbf{p},\omega)  .
\end{align}
As mentioned before that all HF SEs are real quantities, hence $\rmupper{\cal I}{HF}{\alpha}(\mathbf{p})$
contributes only to the commonly defined IPD. Here we collect all contributions from statistical correlation 
in the expression of GIPD
\begin{equation}
 \rmboth{\cal I}{\,sc}{\alpha}(\mathbf{p}) = 
 \rmupper{\cal I}{HF}{\alpha}(\mathbf{p}) - \rmboth{\Delta}{PF}{bf}(\mathbf{p}).
\end{equation}

Inserting Eq.~\eqref{Intint} into Eq.~\eqref{intContrib22} and
introducing the following expression
\begin{align}\label{AAfunc}
 {\cal A}_\alpha (\mathbf{p,k},\omega)  = & V_{\alpha \alpha}(\mathbf{k})\, A_\alpha(\mathbf{p-k},\omega) \nonumber \\ & 
- V_{(\alpha+1)(\alpha+1)}(\mathbf{k})\, A_{\alpha+1}(\mathbf{p-k},\omega) \nonumber \\ &
-V_{\mathrm{ee}}(\mathbf{k})\, A_{\mathrm{e}}(\mathbf{p-k},\omega) 
\end{align}
to describe atomic properties of the ion involved in the ionization reaction,
the dynamical correlation part of the GIPD, i.e. Eq.~\eqref{intContrib22}, can be rewritten as
\begin{equation}\label{corrcontrib1}
 \rmupper{{\cal I}}{corr}{\alpha}\!(\mathbf{p},\omega)\! =\! \!\sum_{\mathbf{k}} \!\int_{-\infty}^{\infty}\! \!
 \frac{d \omega_1}{2\pi} {\cal A}_\alpha (\mathbf{p,k},\omega_1)  M(\mathbf{k},z_\nu,\omega_1) .
\end{equation}
With the expression~\eqref{Mfunction} the real and imaginary part of the dynamical correlation 
contribution~\eqref{corrcontrib1} take the following expressions 
\begin{widetext}
\begin{align}
  \rmboth{\cal I}{\,dc}{\alpha}(\mathbf{p},\omega)  & = \repart \rmupper{\cal I}{corr}{\alpha} (\mathbf{p},\omega) 
  =  {\cal P} \sum_{\mathbf{k}} \int_{-\infty}^{\infty} 
 \frac{d \omega_1}{2\pi} \int_{-\infty}^{\infty}  \frac{d\omega_2}{\pi} \,
 \frac{{\cal A}_\alpha (\mathbf{p,k},\omega_1)}{\omega  - \omega_1 - \omega_2} \cdot 
 \impart\! \! \squarebra{ \frac{ \rmnumb{n}{B}(\omega_2) + 1 }{ \varepsilon(\mathbf{k},\omega_2)} }, 
 \label{realIPD} \\ & \nonumber \\
 \rmboth{\cal B}{\,dc}{\alpha}(\mathbf{p},\omega) & =  \impart \rmupper{\cal I}{corr}{\alpha}\! (\mathbf{p},\omega) 
 = - \sum_{\mathbf{k}} \int_{-\infty}^{\infty} \frac{d \omega_1}{2\pi} \,{\cal A}_\alpha (\mathbf{p,k},\omega_1) \cdot
 \impart\! \! \squarebra{ \frac{ \rmnumb{n}{B}(\omega-\omega_1) + 1 }{ \varepsilon(\mathbf{k},\omega-\omega_1)} } 
 \label{imagIPD} .
\end{align}
\end{widetext}
The essential quantities determining the GIPD are the spectral function $A_c(\mathbf{p},\omega)$
and the dielectric function $\varepsilon(\mathbf{k},\omega)$. The spectral function $A_c(\mathbf{p},\omega)$ 
of particle $c$ will be discussed in detail in the subsequent subsection.
The dielectric function $\varepsilon(\mathbf{k},\omega)$ describes dielectric response of a material 
to an external field. A large amount of physical properties, such as stopping power, optical spectra, and 
conductivity, are directly connected to the dielectric function~\cite{KKER86,KSKB05}. The dielectric function 
is extremely complicate to be determined. A famous approximation for the dielectric function 
$\varepsilon(\mathbf{k},\omega)$ is the random phase approximation~\cite{AB84}, which describes the effective
response in a non-interacting gas.
Improvements based on the random phase approximation can be performed, such as local field correlation~\cite{UI80,FWR10}, 
Born-Mermin ansatz~\cite{Mermin65}, and the extended Born-Mermin ansatz~\cite{RSWR99,SRWRPZ01}. 
Attempts to develop dielectric function for 
two-component plasma have been proposed by different authors~\cite{SRWRPZ01,Roepke98,RW98,AAADT14}.
However, all these improvements are inadequate to describe the effective charge response in a 
multicomponent plasmas under warm dense matter conditions.
In this work we express the dielectric function $\varepsilon(\mathbf{k},\omega)$ in terms of 
the charge-charge dynamical SF $\rmnumb{S}{zz}(\mathbf{k},\omega)$ 
according to the fluctuation-dissipation theorem. Furthermore, the proposed approach in this work can be improved by
taking higher-order cluster contributions, e.g. through considering the multiple interaction in the T-matrix 
approximation~\cite{LRKR17}. Such improvements will not be handled within the context of the present investigation.

\section{Ionization potential depression: statistical and dynamical correlations in many-body systems}
\label{IPDandSF}

In this section we demonstrate that the IPD in a multi-component Coulomb system can be directly related 
to the spatial distribution and temporal fluctuation of the plasma environment, i.e. the dynamical SFs.
The SPSE incorporated in the GIPD includes the HF term and the correlation contribution.
Correspondingly,  the GIPD can be decomposed into two contributions, i.e. contribution from the 
statistical (or exchange) correlation and from the dynamical correlation, respectively.
The dynamical correlation contribution $\rmupper{\Sigma}{corr}{c}$ describes dynamical interactions between 
the investigated system (denoted as impurity) and its surrounding charged environment (treated as perturber). 
The exchange correlation includes the HF contribution of the continuum edge $\rmupper{\Sigma}{HF}{c}$,  
the Pauli blocking $\rmboth{\Delta}{Pauli}{bf}$, and Fock shift for bound states $\rmboth{\Delta}{Fock}{bf}$
stemming from spin statistics of identical particles. In this section we will discuss the statistical and dynamical 
correlations in many-body systems, where the real part of GIPD, i.e. the commonly defined IPD, is of central relevance.
The uncertainty of IPD, characterized by broadening of the continuum edge and energy levels, will be briefly discussed 
in the present study. More details will be presented in a forthcoming work.

\subsection{Approximation for spectral function $A_c(\mathbf{p},\omega)$}
The spectral function $A_c(\mathbf{p},\omega)$ contains all information about the dynamical
behavior of a particle in an interacting many-body environment and satisfies the normalization condition
$\int_{-\infty}^\infty \frac{d\omega}{2\pi} A_c(\mathbf{p},\omega) = 1$. It is related to the SPSE
$\Sigma_c(\mathbf{p},\omega)$ as follows~\cite{KKER86,KSKB05,SL13}
\begin{equation}\label{spectralLF}
 A_c(\mathbf{p},\omega) = \frac{ 2 \Gamma_c(\mathbf{p},\omega) }
 { \squarebra{ \omega - E_c(\mathbf{p}) - \Delta_c(\mathbf{p},\omega) }^2 
 + \squarebra{\Gamma_c(\mathbf{p},\omega)}^2 }.
\end{equation}
As shown in Eq.~\eqref{singleSE}, the SPSE is further connected to the spectral function, so that
the spectral function $A_c(\mathbf{p},\omega) $ and the SPSE $\Sigma_c(\mathbf{p},\omega)$ have to be 
determined self-consistently. In the case of non-interacting gases or in the case of negligible
width ($\Gamma_c \rightarrow 0$) of weakly interacting gases, the Lorentz form of spectral function, 
i.e. Eq.~\eqref{spectralLF}, can be replaced by a simple $\delta$-shape
\begin{equation}
 A_c(\mathbf{p},\omega) =  2\pi \, \delta\bks{\hbar \omega - E_c(\mathbf{p})}.
\end{equation}
Such simplified treatment of the spectral function results in the $G_0 W$ approximation of the SPSE,
where $G_0$ denotes the undressed Green's function for free particles.
Note that we actually do not know which approximation is better for the evaluation of SE.
It is well know in condense matter physics that the self-consistent GW approximation usually results in 
a good quasi-energy but overestimates the energy gap, while $G_0 W$ approximation with screened potential $W$
in RPA level gives a better energy gap~\cite{Holm99,KNO16}. For a consistent approach, vertex corrections have also to
be included in the calculation.

In the following, we perform the $G_0W$ approximation for the SPSE to calculate the IPD in multicomponent plasmas. 
Then we have for the function ${\cal A}(\mathbf{p,k},\omega)$,
i.e. Eq.~\eqref{AAfunc}, in the expression of GIPD~\eqref{corrcontrib1}
\begin{align}
{\cal A}(&\mathbf{p,k}, \omega)
  = -\frac{2\pi\, \bks{z_\alpha+1}^2 e^2}{\varepsilon_0 k^2}\,\delta\!\bks{\hbar \omega - {\cal E}_{\alpha+1,\mathbf{p-k}}} \\
- &  \frac{2\pi\,  e^2}{\varepsilon_0 k^2}\,\delta\!\bks{\hbar \omega - {\cal E}_{\mathrm{e},\mathbf{p-k}}} 
+ \frac{2\pi\, z_\alpha^2 e^2}{\varepsilon_0 k^2}\,\delta\!\bks{\hbar \omega - {\cal E}_{\alpha,\mathbf{p-k}}} .\nonumber
\end{align}
As discussed in Ref.~\cite{LRKR17} (for details also see Appendix~\ref{derivationGIPD}), we take the momentum $\mathbf{p} = 0$ after performing 
classical dispersion relation
$\hbar \omega = {\cal E}_{\alpha,\mathbf{p}}$ for the investigated ion $\alpha$, i.e.
$\rmboth{\cal I}{\text{\tiny GIPD}}{\alpha}(\mathbf{p},\omega) \rightarrow 
\rmboth{\cal I}{\text{\tiny GIPD}}{\alpha}\bks{\mathbf{p},{\cal E}_{\alpha,\mathbf{p}}/\hbar}\rightarrow 
\rmboth{\cal I}{\text{\tiny GIPD}}{\alpha}(0,0)$. In other words, we define the GIPD as
$\rmboth{\cal I}{\text{\tiny GIPD}}{\alpha} =\rmboth{\cal I}{\text{\tiny GIPD}}{\alpha}(0,0) $. 
In order to achieve an analytic expression for the IPD, 
a further simplification in the derivation is to take the classic limit $\hbar \rightarrow 0$ in the propagator
$1/\squarebra{-\omega'-\hbar k^2/(2m_c)}$ in Eq.~\eqref{realIPD} for the real part of the correlation contribution.
Under these approximations we have
\begin{equation}\label{QQ111}
 \int_{-\infty}^\infty \frac{d \omega_1}{2\pi} 
 \frac{{\cal A}_\alpha (\mathbf{p,k},\omega_1)}{\omega  - \omega_1 - \omega_2}
 \approx \frac{2\bks{z_\alpha+1}\, e^2}{\varepsilon_0\, k^2\, \omega_2},
\end{equation}
with which the real part of dynamical interaction contribution in the GIPD, i.e. Eq.~\eqref{realIPD},
is given by
\begin{equation} \label{IPDdielectricReal}
 \rmboth{\cal I}{\, dc}{\alpha} \!\! =\!\! \int \!\!\! \frac{d^3 \mathbf{k}}{\bks{2\pi}^3} \!\!
 \int_{-\infty}^\infty\!\! \frac{d \omega_2}{\pi}
 \frac{2\bks{ \! z_\alpha+1 \! } e^2}{\varepsilon_0\, k^2 \, \omega_2} 
 \impart\! \! \squarebra{ \!\frac{ \rmnumb{n}{B}(\omega_2) + 1 }{ \varepsilon(\mathbf{k},\omega_2)}\! }.
\end{equation}
Similarly, the broadening of the IPD, i.e. Eq.~\eqref{imagIPD}, can be described by
\begin{align}\label{IPDdielectricImag}
 \rmboth{\cal B}{\, dc}{\alpha} & =\int \frac{d^3 \mathbf{k}}{\bks{2\pi}^3}
 \int_{-\infty}^\infty d \omega_2 \, \frac{2\bks{z_\alpha+1}\, e^2}{\varepsilon_0\, k^2 } \nonumber \\
 & \quad \times \impart\! \! \squarebra{ \frac{ \rmnumb{n}{B}(\omega_2) + 1 }{ \varepsilon(\mathbf{k},\omega_2)} }\, 
 \delta\!\bks{\omega_2 + {\cal E}_{\alpha,\mathbf{k}}/\hbar} .
\end{align}
Detailed derivations of Eq.~\eqref{QQ111}, Eq.~\eqref{IPDdielectricReal} and Eq.~\eqref{IPDdielectricImag} 
are given in Appendix~\ref{derivationGIPD}.

\subsection{IPD due to statistical correlation: Hartree-Fock contribution and Pauli blocking} 
In this subsection we discuss the statistical correlation contribution of the GIPD
\begin{equation}
 \rmupper{\cal I}{sc}{\alpha} = \rmupper{\cal I}{HF}{\alpha} - \rmboth{\Delta}{PF}{bf} .
\end{equation}
As described in Sec.~\ref{DefineIPD}, here the Hartree-Fock contribution to the continuum edge 
$\rmupper{\cal I}{HF}{\alpha}$ and the energy shift of bound states $\rmboth{\Delta}{PF}{bf}$ 
due to Pauli blocking $\rmboth{\Delta}{Pauli}{bf}$ and Fock shift $\rmboth{\Delta}{Fock}{bf}$ 
are taken into account.

\subsubsection{Hartree-Fock contribution to the continuum edge}
Within the $G_0W$ approximation, the HF term of the SPSE
contained in the HF contribution in the GIPD, i.e. $ \rmupper{\cal I}{HF}{\alpha} = \rmupper{\Sigma}{HF}{\alpha}
- \rmupper{\Sigma}{HF}{\alpha+1} - \rmupper{\Sigma}{HF}{\mathrm{e}}$, is given by
\begin{equation}\label{HFSPSE}
 \rmupper{\Sigma}{HF}{c}  = -\int \frac{d^3\mathbf{k}}{(2\pi)^3}\, \frac{z_c^2 e^2}{\varepsilon_0\, k^2}\, n_c(\mathbf{k}).
\end{equation}
Inserting the explicit expression for the distribution function $n_c(\mathbf{k})$, i.e.~\eqref{distributionFunc}, into 
Eq.~\eqref{HFSPSE} yields
\begin{align}\label{HFcontrib1}
\rmupper{\Sigma}{HF}{c} 
= \pm\, \frac{z_c^2\, e^2}{2\pi\, \varepsilon_0\, \lambda_c}\cdot \mathrm{Li}_{1/2}\bks{\mp\, e^{\beta \mu_c}},
\end{align}
where the upper/lower sign corresponds to the case of fermions/bosons according to spin statistics 
and $\mathrm{Li}_n(z)$ is the polylogarithm function. For non-degenerate ions, the Fermi-Dirac or
the Bose-Einstein distribution can be approximated by the 
Maxwell-Boltzmann distribution. This approximation yields the following expression of HF SE
\begin{equation}
 \rmupper{\Sigma}{HF,cl}{c} = -\frac{z_c^2\, e^2}{2\pi\, \varepsilon_0\, \lambda_c}\cdot  e^{\beta \mu_c}.
\end{equation}

\begin{figure}[t]
\centering
 \includegraphics[width=0.48\textwidth]{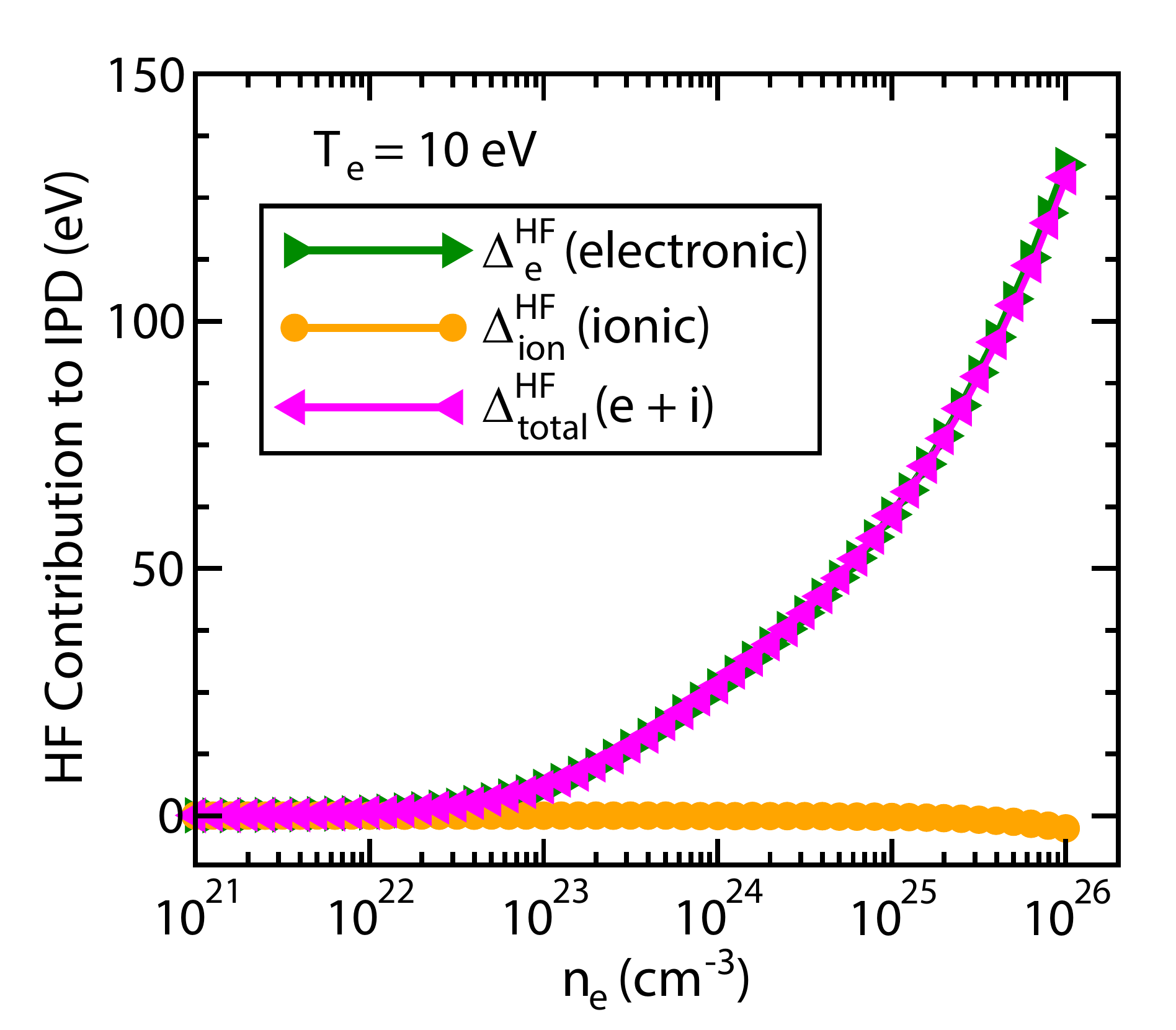}
\caption{HF contribution to the total IPD at a fixed temperature $\rmnumb{T}{e} = 10$ eV with varying free
electron density $\rmnumb{n}{e}$ in a 4-fold ionized carbon plasma. The electronic (green line with triangle right)
and ionic (yellow line with circle) HF contribution
are calculated according Eq.~\eqref{HFcont22} and~\eqref{HFcont11}, respectively. The sum of them yields the total 
HF contribution~\eqref{HFcont00} (magenta line with triangle left). }
\label{HFcontinuum}
\end{figure}

Consequently, the HF contribution to the continuum edge, i.e. 
$\rmupper{\cal I}{HF}{\alpha} = \rmupper{\Sigma}{HF}{\alpha}- 
\bks{\rmupper{\Sigma}{HF}{\alpha+1} + \rmboth{\Sigma}{HF}{e} }$, is given by 
\begin{equation}\label{HFcont00}
 \rmupper{\cal I}{HF}{\alpha}  = \rmboth{\Delta}{HF}{ion} + \rmboth{\Delta}{HF}{e}
\end{equation}
with the electronic contribution
\begin{equation}\label{HFcont22}
 \rmboth{\Delta}{HF}{e} = -\frac{e^2}{2\pi\varepsilon_0 \rmnumb{\lambda}{e}}
 \cdot \mathrm{Li}_{1/2}\bks{- \,e^{\beta \rmnumb{\mu}{e}}},
\end{equation}
and the ionic contribution 
$\rmboth{\Delta}{HF}{ion}  = \rmupper{\Sigma}{HF}{\alpha}-\rmupper{\Sigma}{HF}{\alpha+1}$
\begin{align}\label{HFcont11}
 \rmboth{\Delta}{HF}{ion} = \frac{\bks{z_\alpha+1}^2 e^2}{2\pi\varepsilon_0 \rmnumb{\lambda}{\alpha}}
 \cdot e^{\beta \mu_\alpha}\! 
 \squarebra{e^{-\beta \rmnumb{\mu}{e}}- \bks{\frac{z_\alpha}{z_\alpha+1}}^2  }.
\end{align}
Here the chemical potentials of ideal gases for ions are used. For non-degenerate ions,
$\mu_\alpha = \rmboth{\mu}{id}{\alpha} = \kbt{}\,\,\mathrm{ln}\!\bks{n_\alpha \lambda_\alpha^3/g_\alpha}$ 
is applied, where $g_\alpha$ is the statistical weight of the ground state of ionization state $\alpha$.
For electronic chemical potential $\rmboth{\mu}{id}{e}$, the expression~\eqref{chemEEideal} is utilized in the calculation. 
The results for a $\mathrm{C}^{4+}$ plasma at temperature $\rmnumb{T}{e} = 10$ eV are shown in Fig.~\ref{HFcontinuum}.
It can be seen that the HF SE of the ion $\alpha$ and its next ionization stage $\alpha+1$ compensates with each other,
so that the whole ionic contribution is negligible in comparison to the electronic contribution. Even in the case of
low density for $\rmnumb{n}{e} < 10^{21} \mathrm{cm}^{-3}$, the ionic contribution amounts to $\sim 1\%$ 
of the electronic one. Therefore, in most case only the electronic contribution, i.e. Eq.~\eqref{HFcont22}, 
has to be taken into account for the determination of IPD in plasmas.

\subsubsection{Energy shift of bound states}
The impact of the statistical correlation is not only reflected in the reduction of the continuum edge
$\rmupper{\cal I}{HF}{\alpha}$ but also in the shift of discrete energy levels $\rmboth{\Delta}{PF}{bf} $.
We assume that the optical electron are ionized from the outermost shell of ion $\alpha$.
Therefore, in this section the subscript $\mathrm{(\alpha, e)}$ is used to denote the bound electron 
in the outermost shell of the ground state of charge state $\alpha$. 
The shift $\rmboth{\Delta}{PF}{bf} $ consists of the Fock shift~\cite{Roepke19}
\begin{equation}\label{Fockshift}
 \rmboth{\Delta}{Fock}{bf} = -\sum_{\mathbf{p,k}} \rmboth{\phi}{2}{\alpha,e}\!\bks{\mathbf{p}}
 \rmnumb{f}{e}\!\bks{\mathbf{k}} \frac{e^2}{\varepsilon_0\bks{\mathbf{p-k}}^2}
\end{equation}
and the Pauli blocking shift~\cite{Roepke19}
\begin{equation}\label{Paulishift}
 \rmboth{\Delta}{Pauli}{bf} = \sum_{\mathbf{p,k}} \rmboth{\phi}{*}{\alpha,e}\!\bks{\mathbf{p}}
 \rmboth{\phi}{}{\alpha,e}\!\bks{\mathbf{k}}
 \rmnumb{f}{e}\!\bks{\mathbf{p}}
 \frac{\rmupper{z}{}{\alpha,e}\, e^2}{\varepsilon_0\bks{\mathbf{p-k}}^2}.
\end{equation}
The bound electron is assumed to move in the mean field produced by the nucleus and other bound electrons.
The effective charge number experienced by a bound electron in the 
outermost shell of charge state $\alpha$ in the isolated case is described by $\rmupper{z}{}{\alpha, e}$. 
It can be calculated within the screened hydrogenic model~\cite{FBR08}, where a many-electron atom/ion 
is approximated by a hydrogen-like system. Assuming that the selected bound electron occupies 
the $1s$-like state of the hydrogen-like system with effective core charge number $\rmupper{z}{}{\alpha, e}$,
then the corresponding wave function reads
\begin{equation}
 \rmboth{\phi}{}{\alpha,e}\!\bks{\mathbf{r}} = \bks{\pi\, \rmboth{a}{3}{\alpha,e}}^{-1/2}\,
 e^{-r/\rmnumb{a}{\alpha,e}}
\end{equation}
with $\rmnumb{a}{\alpha,e} = a_0 / \rmupper{z}{}{\alpha, e}$. In the momentum space 
the wave function is given by
\begin{equation}
 \rmboth{\phi}{}{\alpha,e}\!\bks{\mathbf{p}} = \int d^3 \mathbf{r} \, e^{i\mathbf{p\cdot r}}
 \rmboth{\phi}{}{\alpha,e}\!\bks{\mathbf{r}}
 =  \frac{ 8\,\bks{\pi\, \rmboth{a}{3}{\alpha,e}}^{1/2} }{ \bks{1+ p^2\,\rmboth{a}{2}{\alpha,e}  }^2 }.
\end{equation}
Inserting this wave function into the Fock shift~\eqref{Fockshift} as well as into the Pauli blocking term~\eqref{Paulishift}, we obtain
\begin{align}
  \rmboth{\Delta}{Fock}{bf} = & - \frac{e^2\, \rmnumb{a}{\alpha,e}^2 }{6\, \pi^2\, \varepsilon_0}
 \int_0^\infty dp\, \frac{p^2\,  \rmnumb{f}{e}\!\bks{p} }{ \bks{1+ p^2\,\rmboth{a}{2}{\alpha,e}  }^3 } \nonumber \\
 & \times \bks{ 3\, p^4\,\rmboth{a}{4}{\alpha,e} + 10\, p^2\,\rmboth{a}{2}{\alpha,e} + 15 }
\end{align}
and 
\begin{equation}
 \rmboth{\Delta}{Pauli}{bf} 
 =  \frac{4\,\rmnumb{z}{\alpha, e} \, e^2\, \rmnumb{a}{\alpha,e}^2 }{\pi^2\, \varepsilon_0}
 \int_0^\infty dp\, \frac{p^2\,  \rmnumb{f}{e}\!\bks{p} }{ \bks{1+ p^2\,\rmboth{a}{2}{\alpha,e}  }^3 }.
\end{equation}

Gathering both contribution and referring it as Pauli-Fock contribution, we arrive at
\begin{equation}
 \rmboth{\Delta}{PF}{bf} =  \frac{e^2}{ \pi^2\, \varepsilon_0\,a_0}
 \int_0^\infty dp_0\,\, \frac{p_0^2\,  \rmnumb{f}{e}\!\bks{p_0} \, y(p_0) }{ \bks{1+ p_0^2/\rmboth{z}{2}{\alpha,e}  }^3 }
\end{equation}
with $p_0 = p\, a_0$ and
\begin{equation}
 y(p_0) = \frac{4}{\rmnumb{z}{\alpha, e}} 
 - \frac{3\, p_0^4/\rmboth{z}{4}{\alpha,e} + 10\, p^2_0/\rmboth{z}{2}{\alpha,e} + 15}{6\,\rmnumb{z}{\alpha, e}^2 }.
\end{equation}
It can be demonstrated that the bound electrons have an important influence on the physical properties, 
in particular impacted by the Pauli blocking effect in strongly degenerate plasmas~\cite{Roepke19,Lin19}. 
A more detailed description demands a systematic investigation of the internal structure and the knowledge of the interaction between the bound electrons in complex many-electron systems, which is not intended in the present work.

%
%

\subsection{IPD due to dynamical correlation: real part of dynamical interaction contribution}
 In this section we discuss the dynamical interaction contribution of the GIPD, where the charge-charge
 dynamical SF $\rmnumb{S}{zz}(\mathbf{k}, \omega) $ are introduced according to the fluctuation-dissipation 
 theorem to describe the influence of the plasma environment on the investigated atomic/ionic system.  
 The IPD in plasmas is proved to be directly determined by the spatial arrangement and the temporal fluctuation 
 of surrounding particles.

\subsubsection{Fluctuation-dissipation theorem}
The response of an interactive plasma system to external perturbations are totally determined by
the dielectric function, which contains the complete information on the ions and the free electrons in this interacting 
system. It is directly connected to the density-density response function $\chi_{cd}({\mathbf k},\omega)$ 
between particles of species $c$ and $d$ via the following relation~\cite{KSKB05,GR09}
\begin{equation}
 \varepsilon^{-1}({\mathbf k},\omega) = 1 + \frac{1}{\varepsilon_0\, k^2}
 \sum_{cd}e_c\, e_d\, \chi_{cd}({\mathbf k},\omega),
\end{equation}
which describes the induced density fluctuations of species $c$ owing to the influence 
of an external field on particles of species $d$. Additionally, the detailed spatial and temporal structure of 
a many-body system is elaborately described by its density-density dynamical SF. Such dynamical SF
determines many transport and optical properties, such as stopping power, the equation of state, the spectral 
lines,  IPD and the corresponding ionization balance. 
Taking into account the fact that the dynamical SFs are related to their corresponding density-density 
correlation functions $\langle \delta n_c({\mathbf r},t) \delta n_d(0,0) \rangle$ via Fourier transformation, 
the partial density-density dynamical SF for different components in plasmas can be defined in terms of
the density-density response function $\chi_{cd}({\mathbf k},\omega)$ via the following
expression~\cite{Hoell07}
\begin{equation}
 S_{cd}({\mathbf k},\omega) \!=\! \frac{\hbar}{\pi\, \sqrt{n_c n_d} }\, \frac{1}{1-e^{-\hbar\omega/\kbt{}}} \, 
 \impart \chi_{cd}({\mathbf k},\omega)
\end{equation}
Therefore, for a multi-component plasma the fluctuation-dissipation theorem can be described
by means of the following relation
\begin{align}
 \impart \squarebra{ \frac{1+ \rmnumb{n}{B}(\omega)}{\varepsilon({\mathbf k},\omega) } }
 \!=\!  \frac{\pi e^2}{\hbar \varepsilon_0 k^2} 
 \sum_{cd} z_c z_d \sqrt{n_cn_d} \, S_{cd}({\mathbf k},\omega).
\end{align}
The free-bound dynamical SF $ S_{\mathrm{e}\gamma}\bks{\mathbf{k},\omega}$ accounting for the correlation between bound and free electrons and the electron-electron dynamical SF $\rmnumb{S}{ee}\bks{\mathbf{k},\omega}$ are related to 
the ionic dynamical SF $\rmnumb{S}{\gamma\nu}\bks{\mathbf{k},\omega}$ and the
dynamical SF $\rmnumb{S}{ee}^0\bks{\mathbf{k},\omega} $ of fast-moving free electrons~\cite{Chihara00,WVGG11}
\begin{align}
 S_{\mathrm{e}\gamma}\bks{\mathbf{k},\omega} 
 & = \sum_\nu \bks{\frac{x_\nu}{\bar{z}}}^{1/2}\,  q_\nu(\mathbf{k}) S_{\gamma\nu}(\mathbf{k}, \omega), \\
\rmnumb{S}{ee}\bks{\mathbf{k},\omega} & = \sum_{\mu\nu} \frac{q_\mu(\mathbf{k})\, q_\nu(\mathbf{k})}{\bar{z}} \bks{x_\mu x_\nu}^{1/2}\,
 S_{\mu\nu}(\mathbf{k}, \omega) \nonumber \\ & \qquad + \rmnumb{S}{ee}^0\bks{\mathbf{k},\omega} .
\end{align}
Within the framework of the linear response theory the screening function $q_\gamma(\mathbf{k})$ 
can be expressed in terms of the electronic dielectric function~\cite{Chihara00,GVWG10,Chapman15}
\begin{equation}
 q_\gamma(\mathbf{k}) = V_{\mathrm{e}\gamma}(k)\, \frac{ 1 - \rmnumb{\varepsilon}{ee}(k,0) }
 { \rmnumb{V}{ee}(k)\,\rmnumb{\varepsilon}{ee}(k,0) } .
\end{equation}
Taking  electron-ion  interactions  to  be  Coulomb interaction potentials, the long-wavelength limit of the screening function within the linear response is given by~\cite{GVWG10,Chapman15}
\begin{equation}\label{qscLW}
 q_\gamma(\mathbf{k}) = z_\gamma \, \frac{\rmnumb{\kappa}{e}^2}{k^2+\rmnumb{\kappa}{e}^2},
\end{equation}
which is proportional to the charge number $z_\gamma$ of  the test particle $\gamma$. 
In the high density regime, such long-wavelength approximation hidden in the RPA dielectric function 
might be inapplicable to describe the finite-wavelength screening, so that the full version of RPA dielectric function
has to be utilized in the calculation of the screening function $q_\gamma(\mathbf{k}) $~\cite{Chapman15}. 

Consequently, the fluctuation-dissipation theorem can be recast in terms of the
total charge-charge dynamical SF $\rmnumb{S}{zz}(\mathbf{k}, \omega) $.
The effective charge-charge response of a multi-component charged plasma to an immersed 
impurity is then described by
\begin{align}\label{fdtCharge}
 \impart \squarebra{ \frac{1+ \rmnumb{n}{B}(\omega)}{\varepsilon({\mathbf k},\omega) } }
=  \frac{\pi \kbt{}}{\hbar \, k^2}\, \rmnumb{\kappa}{scr}^2 \,  
\rmnumb{S}{zz}(\mathbf{k}, \omega) 
\end{align}
with respect to the total inverse screening length $\rmnumb{\kappa}{scr}$. Detailed explanation 
of this expression is given in the Appendix~\ref{CCDSF}. The inverse screening parameter $\rmnumb{\kappa}{scr}$ is 
exhaustively discussed in the next subsection (also see the Appendix~\ref{screeningTheory}). 
The total charge-charge dynamical SF is given by
\begin{align}
\rmnumb{S}{zz}(\mathbf{k}, \omega) 
& = \frac{\rmnumb{S}{ee}^0\bks{\mathbf{k},\omega} }{1+\rmnumb{z}{p}}  + 
\sum_{\mu\nu} \frac{z_\mu z_\nu}{\bar{z}\bks{1+ \rmnumb{z}{p}}}\,  \bks{x_\mu x_\nu}^{1/2} \\ & \quad \times
\bks{ 1 - \frac{ q_\mu(\mathbf{k})}{z_\mu} } \bks{ 1 - \frac{ q_\nu(\mathbf{k})}{z_\nu} }
 S_{\mu\nu}(\mathbf{k}, \omega) \nonumber .
\end{align}
For the screening function $q_\gamma(\mathbf{k})$, the long-wavelength approximation, i.e. Eq.~\eqref{qscLW}, will be used in the following calculations.
Introducing
\begin{equation}\label{screeningLRT}
 \rmnumb{q}{scr}(k) = \frac{q_\gamma(k)}{z_\gamma} 
 =  \frac{\rmnumb{\kappa}{e}^2}{k^2+\rmnumb{\kappa}{e}^2},
\end{equation}
the effective charge-charge dynamical SF for the total many-body system can be then rewritten as
\begin{equation}\label{CCresponse}
 \rmnumb{S}{zz}(\mathbf{k}, \omega)  =  \frac{1 }{1+\rmnumb{z}{p}} \rmnumb{S}{ee}^0\bks{\mathbf{k},\omega}
 + \frac{\rmnumb{z}{p}}{1+\rmnumb{z}{p}} \rmboth{S}{ion}{zz}(\mathbf{k}, \omega)
\end{equation}
with the ionic charge-charge dynamical SF
\begin{equation}\label{SFionpart}
 \rmboth{S}{ion}{zz}(\mathbf{k}, \omega)\! = \!\big[1 \!- \!\rmnumb{q}{scr}(k)\big]^2 
 \sum_{\mu\nu}\frac{z_\mu z_\nu  \sqrt{x_\mu x_\nu}}{{\bar z}\, \rmnumb{z}{p} } S_{\mu\nu}(\mathbf{k}, \omega) .
\end{equation}

\subsubsection{Non-linear screening effect: effective inverse screening length $\rmnumb{\kappa}{eff}$}

Assuming that the electrostatic potential $\psi_z(r)$ takes the following form
\begin{equation}\label{potDebye}
 \psi_z(r) = \frac{z e}{4\pi\varepsilon_0 \, r} \, \mathrm{exp }\bks{- \rmnumb{\kappa}{scr} r},
\end{equation}
according to the Boltzmann distribution~\cite{KKER86}, the mean particle density of species $j$ in 
the vicinity of the test particle with charge number $z$ is described by 
\begin{equation}
 n_{zj}(r) 
= n_j\, \mathrm{exp } \bks{ - \frac{z_j e\, \psi_z(r)}{\kbt{}} }.
\end{equation}
In a classical plasma,  the inverse screening parameter 
$\rmnumb{\kappa}{scr} = \sqrt{  \rmnumb{\kappa}{e}^2 + \rmnumb{\kappa}{i}^2 }$ 
in Eq.~\eqref{potDebye} is reasonably described by the inverse Debye length with 
$\rmnumb{\kappa}{e}^2 = \rmnumb{n}{e}e^2/(\varepsilon_0\kbt{})$ and $\rmnumb{\kappa}{i}^2 
= \sum_a z_a^2 e^2 n_a/(\varepsilon_0\kbt{}) = \rmnumb{z}{p}e^2\rmnumb{n}{e}/(\varepsilon_0\kbt{})$. 
In a strongly coupled, non-ideal plasma, the electronic screening is determined by the Thomas-Fermi length. 
Similarly, the ionic screening
in high density plasmas can not be described by the Debye length any more, since non-linear effect relating to 
strong ionic coupling has to be taken into account.
In the following, we determine the effective screening parameter $\rmnumb{\kappa}{scr}$ according 
to the perfect screening rule (charge neutrality)~\cite{Nordholm84}
\begin{equation}\label{scrRule}
\int d^3 \mathbf{r}\, \rmnumb{\rho}{scr} (r) = 4\pi \int_0^\infty dr\, r^2 \rmnumb{\rho}{scr} (r)  = - z e,
\end{equation}
where $z$ indicates the charge number of a particular ion after the ionization process has taken place,
i.e. $z = z_\alpha + 1$ is the spectroscopic symbol. The screening cloud is assumed to be spherically 
symmetric and is defined as
\begin{align}
\rmnumb{\rho}{scr} (r)
= \sum_{\nu} (z_\nu e)\, n_{z  \nu}(r) - e\, n_{z\mathrm{ e}}(r) .
\end{align}
Inserting this expression into the Eq.~\eqref{scrRule}, an effective inverse screening parameter 
$\rmnumb{\kappa}{scr}$ can be determined,
if the detailed charge state distribution is known. However, such procedure is very complicated to be performed, since
integration over a series of transcendental functions has to be worked out. Regarding the ionic mixture effectively as a single ionic species with charge $\rmnumb{z}{p}e$, the screening cloud can be approximated by the following expression
\begin{equation}
 \rmnumb{\rho}{scr}  (r\!)  \!
 \approx \! e \rmnumb{n}{e} \! \curlybra{\! \mathrm{exp}\!\!
 \squarebra{\! - \frac{\bks{z_\alpha \!+ \! 1}  \bks{\rmnumb{z}{p} \!+ \! 1} e^2}
 {4\pi \varepsilon_0 r \kbt{} }\, e^{-r\rmnumb{\kappa}{scr}} \! } \!\!-\!1\! }.
\end{equation}
The derivation of this approximation is given in detail in Appendix~\ref{screeningTheory}.
Introducing the following dimensionless variables
\begin{align}
x_\alpha  = r / r_\alpha , \quad \rmnumb{\kappa}{eff} = \rmnumb{\kappa}{scr} \, r_\alpha  , \nonumber 
\end{align}
and the impurity-perturber coupling strength
\begin{align}
 \Gamma_\alpha & = \frac{\bks{z_\alpha+1} \bks{\rmnumb{z}{p}+1}\, e^2}{4\pi \varepsilon_0\, r_\alpha\, \kbt{} } 
\end{align}
with respect to the ionic radius
\begin{equation}
 r_\alpha = \bks{\frac{ 3 z }{4\pi \rmnumb{n}{e}}}^{1/3}
= \squarebra{\frac{ 3 \bks{z_\alpha+1} }{4\pi \rmnumb{n}{e}}}^{1/3},
\end{equation}
we obtain the following closed equation for the effective screening parameter $\rmnumb{\kappa}{eff}$
in terms of the impurity-perturber coupling strength $\Gamma_\alpha$
\begin{equation}\label{screeningEffect}
\int_0^\infty\! dx_\alpha \, x_\alpha^2 
\curlybra{ 1 - \mathrm{exp }\! \squarebra{ - \frac{\Gamma_\alpha}{x_\alpha}\, 
\mathrm{exp}\bks{- \rmnumb{\kappa}{eff} \, x_\alpha} } } 
= \frac{1}{3}.
\end{equation}
In a certain plasma condition with given density $\rmnumb{n}{e}$ and temperature $T$ (corresponding to a fixed coupling strength $\Gamma_\alpha$), the screening parameter $\rmnumb{\kappa}{eff}$ can be determined by solving
the integral equation numerically. For further application, we fit the numerical solutions with following expression
\begin{equation}\label{screening00}
 \rmnumb{\kappa}{eff}^2 
 = \frac{3 \Gamma_\alpha}{ \sqrt{ 1-0.4\,\bks{ 3 \Gamma_\alpha\, \gamma_0^2 }^{3/4} + 3 \Gamma_\alpha\,\gamma_0^2 }}
\end{equation}
with $\gamma_0 = \squarebra{4/(9\pi)}^{1/3}$. In low-density cases within the validity of linear Debye theory, the 
effective inverse screening parameter can be determined analytically via $\rmnumb{\kappa}{eff}^2 \rightarrow r_\alpha^2\, \rmnumb{\kappa}{scr,Debye}^2  = 3 \Gamma_\alpha$.
This can be clearly shown by taking the inverse Debye length for $\rmnumb{\kappa}{e}$ and $\rmnumb{\kappa}{i}$
in the screening parameter $\rmnumb{\kappa}{scr}$. In our previous work~\cite{LRKR17}, the effective 
screening length is approximated by
\begin{equation}\label{screening11}
 \rmnumb{\kappa}{eff,0}^2 
 = \frac{3 \Gamma_\alpha}{ \bks{ 1 + 3 \Gamma_\alpha\,\gamma_0^2 }^{1/2}},
\end{equation}
where the interionic coupling parameter $\rmnumb{\Gamma}{ii}$ is replaced by the impurity-perturber 
coupling strength $\Gamma_\alpha$. Within the framework of the one-component plasma model, 
these two coupling strengths are demonstrated to be equivalent.
In this work, we extend the developed theory in Ref.~\cite{LRKR17} to describe multi-component plasmas, where the 
impurity-perturber coupling strength $\Gamma_\alpha$ is more reasonable to describe the coupling between a relevant
system (regarded as impurity) and its surrounding environment.

\begin{figure}[t]
\centering
 \includegraphics[width=0.48\textwidth]{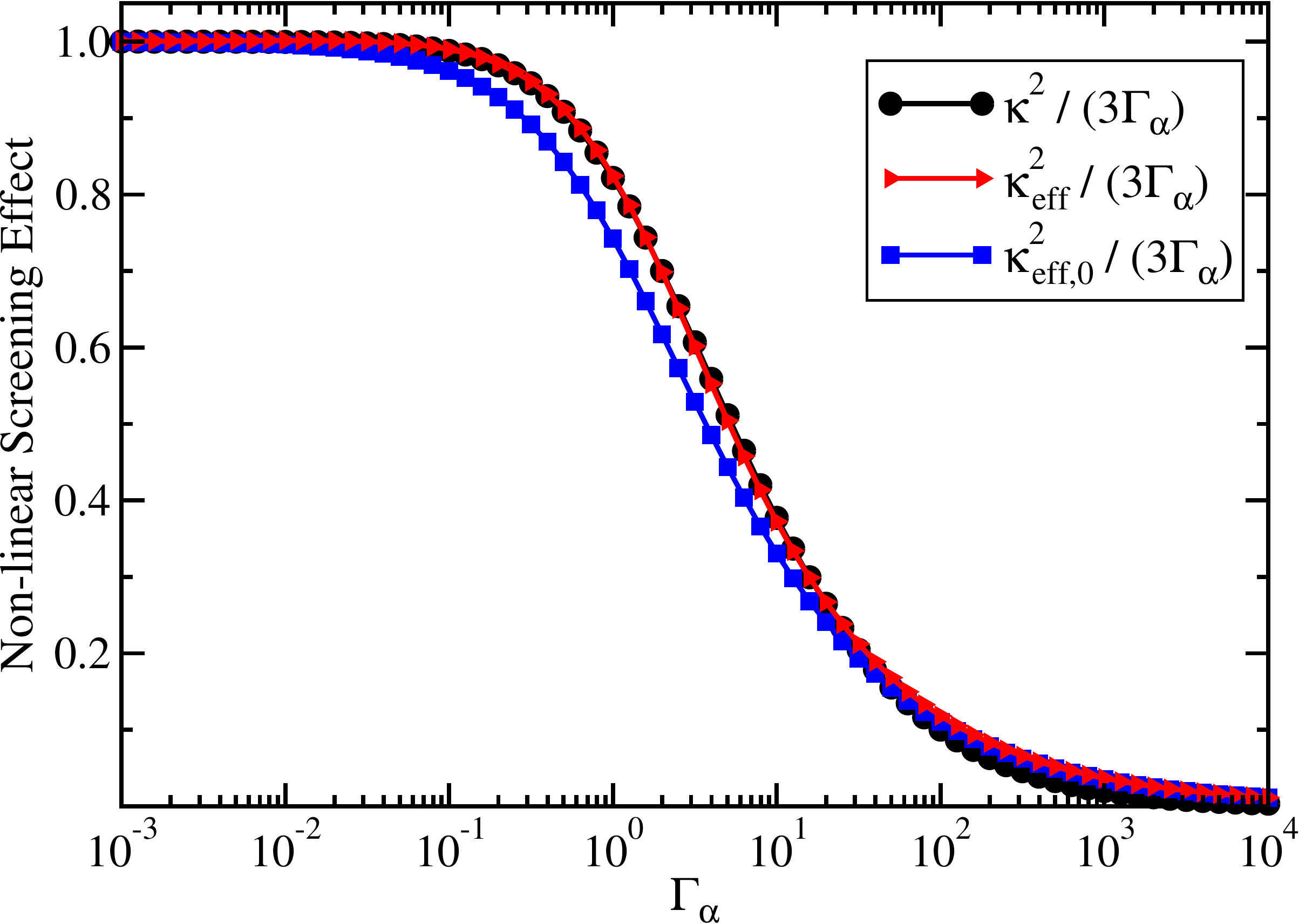}
\caption{Non-linear screening effect due to strong correlations. The black line with circle indicates numerical
solutions of Eq.~\eqref{screeningEffect}. The red line with triangle and the blue line with square denote 
the approximated expressions~\eqref{screening00} and~\eqref{screening11}, respectively.}
\label{nonlinear}
\end{figure}

Within the framework of linear Debye theory, as already discussed, we have 
$\rmnumb{\kappa}{eff}^2/(3 \Gamma_\alpha) \equiv 1$.
Deviation from the linear screening effect is dipicted in Fig.~\ref{nonlinear}. It can be seen that
the linear screening theory is only valid up to $\Gamma_\alpha \approx 0.2$. As the coupling strength
$\Gamma_\alpha$ increases, the difference from the linear screening theory becomes stronger.
Additionally, the solution of Eq.~\eqref{screeningEffect} can be perfectly reproduced 
by the fit formula~\eqref{screening00} in a wide range from weakly coupled tranditional plasma
to crystallization of the plasma ($\Gamma_\alpha \approx 170$).

\subsubsection{Decomposition of the correlation contribution $\rmboth{\cal I}{dc}{\alpha} $}
After the introduction of the fluctuation-dissipation theorem and the discussion on the nonlinear screening 
due to strong coupling effect in many-body systems, we can now express the IPD in terms of the 
dynamical SFs. Inserting the expression~\eqref{fdtCharge} combined with the effective inverse screening parameter
$\rmnumb{\kappa}{eff}$ given by Eq.~\eqref{screening00} into the expression for IPD, 
i.e. Eq.~\eqref{IPDdielectricReal}, yields
\begin{align}
 \rmboth{\cal I}{dc}{\alpha} 
 = \frac{\bks{z_\alpha+1}\, e^2\, \rmnumb{\kappa}{eff}^2 \, a_0}{2\pi^2 \varepsilon_0 \, r_\alpha^2} 
 \int_0^\infty \frac{d k_0}{k_0^2}\, \rmnumb{S}{zz}(k_0)
\end{align}
with the reduced charge-charge SF
\begin{equation}
 \rmnumb{S}{zz}(k_0) = 2\int_{-\infty}^\infty d \omega \, \frac{\kbt{}}{\hbar \omega}\rmnumb{S}{zz}(k_0,\omega),
\end{equation}
where the dimensionless wave number $k_0$ is given by $k_0 = k\, a_0$ with the Bohr radius $a_0$.
According to the fact that the total charge-charge dynamical SF can be decomposed into an electronic
and an ionic contribution, see Eq.~\eqref{CCresponse}, the reduced charge-charge SF can be also split as follows
\begin{equation}
 \rmnumb{S}{zz}(k_0) =  \frac{\rmnumb{z}{p}}{1+\rmnumb{z}{p}} \rmboth{S}{ion}{zz}(k_0)  
 +  \frac{1}{1+\rmnumb{z}{p}} \rmboth{S}{el}{zz}(k_0) ,
\end{equation}
with the reduced ionic charge-charge SF
\begin{align}\label{ionicDSF} 
 \rmboth{S}{ion}{zz}(k_0)  & = 2\int_{-\infty}^\infty d \omega \, 
 \frac{\kbt{}}{\hbar \omega}\, \rmboth{S}{ion}{zz}(k_0,\omega) 
 \end{align}
and the reduced electronic charge-charge SF
\begin{align} \label{eeDSF}
 \rmboth{S}{el}{zz}(k_0) & = 2 \int_{-\infty}^\infty d \omega \, 
 \frac{\kbt{}}{\hbar \omega}\, \rmboth{S}{0}{ee}(k_0,\omega) .
\end{align}
Correspondingly, the dynamical interaction contribution of GIPD $\rmboth{\cal I}{dc}{\alpha} $ can be  separated as follows
\begin{equation}
  \rmboth{\cal I}{dc}{\alpha} = \frac{\rmnumb{z}{p}}{1+\rmnumb{z}{p}} \rmboth{\cal I}{dc,ion}{\alpha} 
  +  \frac{1}{1+\rmnumb{z}{p}} \rmboth{\cal I}{dc,el}{\alpha} 
\end{equation}
with the ionic part of the interaction contribution
\begin{align}
 \rmboth{\cal I}{dc,ion}{\alpha} 
 = \frac{\bks{z_\alpha+1}\, e^2\, \rmnumb{\kappa}{eff}^2 \, a_0}{2\pi^2 \varepsilon_0 \, r_\alpha^2} 
 \int_0^\infty \frac{d k_0}{k_0^2}\, \rmboth{S}{ion}{zz}(k_0), \label{ionicIPD}
 \end{align}
and the electronic part of the interaction contribution
\begin{align}
 \rmboth{\cal I}{dc,el}{\alpha} 
  = \frac{\bks{z_\alpha+1}\, e^2\, \rmnumb{\kappa}{eff}^2 \, a_0}{2\pi^2 \varepsilon_0 \, r_\alpha^2} 
 \int_0^\infty \frac{d k_0}{k_0^2}\, \rmboth{S}{el}{zz}(k_0)  \label{eeIPD} .
\end{align}
In the subsequent two subsections, we will discuss the details about the evaluation of these contributions using 
ionic and electronic SFs in different approximations.

\subsubsection{Ionic part of the correlation contribution to IPD}
The dynamical SF $S\bks{\mathbf{k},\omega}$ is a measurable quantity in scattering experiments,
for example by X-ray Thomson scattering in research field of warm dense matter 
or by neutron-scattering experiments in nuclear physics. In scattering experiments, 
the spectrum for dynamical SFs as a function of the frequency $\omega$ is usually measured at a given 
wavenumber $\mathbf{k}$ (which corresponds to a given scattering angle)~\cite{GR09}. 
In the theoretical modeling, the dynamical SF can be evaluated within the framework of linear response 
theory or by some numerical simulation methods such as molecular dynamics and Monto-Carlo simulation. 
However, these simulation methods are generally restricted due to their large computational
cost. Moreover, in our theory for IPD we have to integrate over the frequency $\omega$ and the 
wavenumber $\mathbf{k}$, so that detailed informations of the dynamical SFs in the whole range of the frequency-wavenumber plane are always indispensable. Such demand can not be easily accomplished with the current 
computing power. Furthermore, in order to obtain the charge state distribution, the coupled Saha equations have
to be solved iteratively in combination with the IPD values. Therefore, some approximations for the dynamical SF
are needed for the calculation of IPD and the determination of ionic fraction in plasmas. In this work the plasmon-pole approximation is applied. In comparison to our previous work~\cite{LRKR17}, the plasmon-pole approximation 
is adapted according to Refs.~\cite{GRHGR07,SRa05,GGa09}
\begin{equation}\label{plasmonpole}
\rmboth{S}{}{\mu\nu}(k,\omega) = \frac{{\cal C}(\omega) }{2}\,\rmboth{S}{}{\mu\nu}(k)\,
\squarebra{\delta(\omega - \rmnumb{\omega}{\mathbf{k}} )+\delta(\omega + \rmnumb{\omega}{\mathbf{k}} )}
\end{equation}
with the prefactor ${\cal C}(\omega) = \hbar \omega \beta/\squarebra{1-\mathrm{exp}\!\bks{-\hbar \omega \beta}}$.
It accounts for the principle of detailed balance 
$\rmboth{S}{}{\mu\nu}\!\bks{\mathbf{k},\omega}/\rmboth{S}{}{\mu\nu}\!\bks{\mathbf{k},\!-\omega} \!= \!\mathrm{exp}\!\bks{\beta \hbar \omega}$~\cite{SRa05}.
The $\mathbf{k}$-dependent frequency $\rmnumb{\omega}{\mathbf{k}}$ is determined by the dispersion relation for ionic acoustic modes in plasmas, which is given
by the relation $\rmnumb{\omega}{\!\mu\nu}^2\!\!\bks{\mathbf{k}}\!\! =\! k^2 \kbt{}/[2M_{\!\mu\nu}\!\bks{1\!-\!\rmnumb{q}{scr}\!(k)}^2\!|\rmboth{S}{}{\!\mu\nu}(k)|]$ in the long-wavelength limit~\cite{HM06} with the reduced ion mass $M_{\mu\nu} = m_\mu m_\nu/(m_\mu+ m_\nu)$.
Consequently, the ionic part of the charge-charge dynamical SF can be expressed in terms of a static one. 
Inserting Eq.~\eqref{SFionpart} in combination with the plasmon-pole ansatz~\eqref{plasmonpole} into 
the reduced ionic charge-charge SF~\eqref{ionicDSF}, we have
\begin{align}
\rmboth{S}{ion}{zz}(k_0)  =  \Big[1-\rmnumb{q}{scr}(k_0)\Big]^2  \rmboth{S}{MIM}{ion}(k_0)  
\end{align}
with the static SF for the multi-ionic mixture 
\begin{equation}
 \rmboth{S}{MIM}{ion}(k_0) 
 = \sum_{\mu\nu}\frac{z_\mu z_\nu  \sqrt{x_\mu x_\nu}}{{\bar z}\, \rmnumb{z}{p} }\, S_{\mu\nu}(k_0) .
\end{equation}
The physical meaning of this plasmon-pole approximation
corresponds to assume that the ions have a fixed distribution in plasmas by neglecting temporal fluctuations. In other
words, because of their large masses compared to free electrons, ionic dynamics are ignored in determining thermodynamical properties. Obviously, the final expression for the ionic correlation contribution is expressed by
\begin{align}
 \rmboth{\cal I}{dc,ion}{\alpha}  
 &= \frac{\bks{z_\alpha+1}\, e^2\, \rmnumb{\kappa}{eff}^2 \, a_0}{2\pi^2 \varepsilon_0 \, r_\alpha^2}
 \int_0^\infty \frac{d k_0}{k_0^2} \Big[1-\rmnumb{q}{scr}(k_0)\Big]^2   \nonumber \\
& \quad \times \sum_{\mu\nu}\frac{z_\mu z_\nu  \sqrt{x_\mu x_\nu}}{{\bar z}\, \rmnumb{z}{p} }\, 
S_{\mu\nu}(k_0) .
\end{align}
The static version of the partial density-density SF $S_{\mu\nu}(k_0)$
can be obtained via different approaches, for example, by solving the Ornstein-Zernike equation 
with the closure relation of Percus-Yevick approximation~\cite{Wertheim63}
or with the hypernetted-chain (HNC) equation~\cite{WHSG08}.
Other numerical simulation methods have also been worked out, such as the density functional theory molecular 
dynamics simulation~\cite{Plagemann12,DSM18} and the path integral Monte-Carlo
simulation~\cite{DSM18}.
In this work, the HNC approach for the ionic static SFs is utilized in the calculation of IPD values and 
the charge state distributions in the Sec.~\ref{resultsDiss}.

\subsubsection{Electronic part of the correlation contribution to IPD}
In this subsection we at first treat electrons and ions at the same footnoting, which should 
be a reasonable approximation for non-degenerate plasmas. Then the plasmon-pole
approximation~\eqref{plasmonpole} for the dynamical SF of free electrons can be performed 
and following expression is obtained for the electronic contribution 
\begin{equation}\label{eeIPDfinal}
 \rmboth{\cal I}{dc,el}{\alpha}  
 = \frac{\bks{z_\alpha+1}\, e^2\, \rmnumb{\kappa}{eff}^2 \, a_0}{2\pi^2 \varepsilon_0 \, r_\alpha^2} 
 \int_0^\infty \frac{d k_0}{k_0^2}\,\rmboth{S}{el}{zz}(k_0)
\end{equation}
with the static SF for free electrons~\cite{GRHGR07}
\begin{equation}\label{eeSSF0}
 \rmboth{S}{el}{zz}(k_0) =
 \rmboth{S}{st}{ee}(k_0) = \frac{k_0^2}{k_0^2 + \bks{\rmnumb{\kappa}{e} \, a_0}^2}.
\end{equation}
In a degenerate plasma, quantum effects and dynamical effect are of essentially importance and have to be 
taken into account for the electronic contribution. To investigate the quantum and dynamic effect of 
free electrons, we have to return to the expression~\eqref{eeDSF}. For this case, a result is obtained by
replacing the static SF~\eqref{eeSSF0} in Eq.~\eqref{eeIPDfinal} by the following expression
\begin{equation}
 \rmboth{S}{el}{zz}(k_0) =\rmboth{S}{dyn}{ee}(k_0) =   2 \int_{-\infty}^\infty d \omega \, 
 \frac{\rmboth{S}{0}{ee}(k_0,\omega)}{\hbar \omega \beta}.
\end{equation}
The dynamical SF in RPA~\cite{AB84,Plagemann12} is utilized in this work.
The static SF~\eqref{eeSSF0} and also the dynamical SF $\rmboth{S}{0}{ee}(k_0,\omega)$ can be improved,
for example, with the local field correlation. Further details about the improvements of  the electronic 
static/dynamical SF can be found in Refs.~\cite{GRHGR07,FWR10,Plagemann12} and the references therein.

%

\subsection{Collection of important formulas }
We conclude this theoretical section with a summary of important formulas for the evaluation of GIPD.
In a many-body environment, the effective ionization potential of a test particle is bounded in the range of 
$\bks{ I_\alpha - \rmupper{\cal I}{sc}{\alpha}- \rmupper{\cal I}{dc}{\alpha} - \frac{1}{2} \rmupper{\cal B}{dc}{\alpha}\ ,
\ I_\alpha - \rmupper{\cal I}{sc}{\alpha} -\rmupper{\cal I}{dc}{\alpha} + \frac{1}{2} \rmupper{\cal B}{dc}{\alpha}}$
with the consideration of the uncertainty of GIPD. 
The effective ionization potential $\rmboth{\cal I}{eff}{\alpha}= I_\alpha - \rmboth{\cal I}{IPD}{\alpha} $
in coupled Saha equations is given in terms of the IPD 
\begin{equation}
 \rmboth{\cal I}{IPD}{\alpha} =  \rmupper{\cal I}{sc}{\alpha}+\rmupper{\cal I}{dc}{\alpha}.
\end{equation}
The contribution $ \rmupper{\cal I}{sc}{\alpha}$ induced by statistical correlations is given by
\begin{equation}\label{ipdSCfinal}
 \rmupper{\cal I}{sc}{\alpha} = \rmboth{\cal I}{HF}{\alpha} - \rmboth{\Delta}{PF}{bf}
\end{equation}
with the Hartree-Fock term from the continuum edge
\begin{equation}
 \rmboth{\cal I}{HF}{\alpha} = -\frac{e^2}{2\pi\varepsilon_0 \rmnumb{\lambda}{e}}
 \cdot \mathrm{Li}_{1/2}\bks{- \,e^{\beta \rmnumb{\mu}{e}}}
\end{equation}
and the Pauli-Fock term from the bound-free coupling
\begin{equation}
 \rmboth{\Delta}{PF}{bf} =  \frac{e^2}{ \pi^2\, \varepsilon_0\,a_0}
 \int_0^\infty dp_0\,\, \frac{p_0^2\,  \rmnumb{f}{e}\!\bks{p_0} \, y(p_0) }
 { \bks{1+ p_0^2/\rmboth{z}{2}{\alpha,e}  }^3 },
\end{equation}
where the parameter function $y(p_0)$ reads
\begin{equation}
 y(p_0) = \frac{4}{\rmnumb{z}{\alpha, e}} 
 - \frac{3\, p_0^4/\rmboth{z}{4}{\alpha,e} + 10\, p^2_0/\rmboth{z}{2}{\alpha,e} + 15}{6\,\rmnumb{z}{\alpha, e}^2 }.
\end{equation}
The interaction contribution $ \rmupper{\cal I}{dc}{\alpha}$ due to the dynamical correlations is described by
\begin{equation}\label{ipdINTfinal}
  \rmboth{\cal I}{dc}{\alpha} = \frac{\rmnumb{z}{p}}{1+\rmnumb{z}{p}} \rmboth{\cal I}{dc,ion}{\alpha} 
  +  \frac{1}{1+\rmnumb{z}{p}} \rmboth{\cal I}{dc,el}{\alpha} ,
\end{equation}
where the ionic part reads
\begin{equation}
 \rmboth{\cal I}{dc,ion}{\alpha}  
 = \frac{\bks{z_\alpha+1}\, e^2\, \rmnumb{\kappa}{eff}^2 \, a_0}{2\pi^2 \varepsilon_0 \, r_\alpha^2} 
 \int_0^\infty \frac{d k_0}{k_0^2}\,\rmboth{S}{ion}{zz}(k_0) 
\end{equation}
with the effective ionic static SF
\begin{equation}
 \rmboth{S}{ion}{zz}(k_0) \!= \!\Big[1-\rmnumb{q}{scr}(k)\Big]^2 \!
 \sum_{\mu\nu}\frac{z_\mu z_nu  \sqrt{x_\mu x_\nu}}{{\bar z}\, \rmnumb{z}{p} }\, S_{\mu\nu}(k_0).
\end{equation}
The electronic part takes the form
\begin{equation}\label{eecontrib}
 \rmboth{\cal I}{dc,el}{\alpha}  
 = \frac{\bks{z_\alpha+1}\, e^2\, \rmnumb{\kappa}{eff}^2 \, a_0}{2\pi^2 \varepsilon_0 \, r_\alpha^2} 
 \int_0^\infty \frac{d k_0}{k_0^2}\,\rmboth{S}{}{ee}(k_0) 
\end{equation}
either with the static SF
\begin{equation}
\rmboth{S}{el}{zz}(k_0)  \rightarrow 
 \rmboth{S}{st}{ee}(k_0) = \frac{k_0^2}{k_0^2 + \bks{\rmnumb{\kappa}{e} \, a_0}^2}\, ,
\end{equation}
or with the reduced dynamical one
\begin{equation}
 \rmboth{S}{el}{zz}(k_0)\rightarrow 
 \rmboth{S}{dyn}{ee}(k_0)  = 2 \int_{-\infty}^\infty d \omega \, 
 \frac{\rmboth{S}{0}{ee}(k_0,\omega)}{\hbar \omega \beta} \, .
\end{equation}


\section{Chemical picture and coupled Saha equations}
\label{CPandSE}
After introducing the effective ionization potential for ions/atoms in an interacting plasma
environment, we can now discuss the ionization equilibrium and the charge state distribution.
In the chemical picture, the composition of a multicomponent plasma can be determined in terms of
a set of coupled Saha equations. For a global ionization/recombination process 
$A_\alpha \rightleftharpoons A_{\alpha+1} + e$, the chemical potentials
of particles involved in the reaction process fulfil the following relation
\begin{equation}\label{ionizationbalance}
 \mu_\alpha = \mu_{\alpha +1} + \rmnumb{\mu}{e}.
\end{equation}
The chemical potentials $\mu_c$ are usually split into an ideal and
an interaction part according to~\cite{KKER86,KSKB05} 
\begin{equation}\label{chemicalCC}
 \mu_c = \rmupper{\mu}{id}{c} + \rmupper{\mu}{int}{c}.
\end{equation}
The first term is the ideal contribution under assumption of non-interacting gas and can
be obtained via the complete Fermi integral $F_{j}(\beta \rmupper{\mu}{id}{c})$ of order $j=1/2$
for arbitrary degeneracy at a given number density $n_c$. The second term accounts for 
the interaction contribution (i.e. medium effects) and results in a modification of the 
ionization potential in a many-body environment.
For free electrons in plasmas, the ideal part is determined from the following normalization
condition for a given electron density~\cite{KSKB05}
\begin{equation}\label{chemEEideal}
 \frac{\rmnumb{n}{e}\, \rmboth{\lambda}{3}{e}}{2}
 = F_{1/2}(\beta \rmupper{\mu}{id}{c})
\end{equation}
with the thermal wavelength $\rmnumb{\lambda}{e} = \sqrt{2\pi \hbar^2/(\rmnumb{m}{e} \kbt{})}$
and inverse temperature $\beta = 1/\bks{\kbt{}}$.
The interaction contribution $\rmboth{\mu}{int}{e}$ can be related to the real part 
of electronic SE, i.e. $\rmboth{\mu}{int}{e} = \repart \rmnumb{\Sigma}{e}$ for free electrons.
However, in the ionization reaction, possibly occupied final states for the ionized electron are
restricted due to the Pauli blocking effect. As already discussed in the Sec.~\ref{DefineIPD}, we define 
the interaction part of the chemical potential for an electron involved in the ionization process as 
(c.f. Eq.~\eqref{ionEEalpha})
\begin{equation}
 \rmboth{\mu}{int}{e} = \repart \rmboth{\Sigma}{e}{ionized}
 = \rmboth{\Delta}{e}{ionized}.
\end{equation}

Instead of performing the separation~\eqref{chemicalCC}, the chemical potential for ions is treated differently
and is defined via the following relation~\cite{SRR09}
\begin{equation}
 \mathrm{exp}\bks{\frac{\mu_\gamma}{\kbt{}}} = \frac{n_\gamma\, \lambda_\gamma^3}{\rmupper{\sigma}{in}{\gamma}}
\end{equation}
with the number density $n_\gamma$ and the thermal wavelength $\lambda_\gamma = \sqrt{2\pi \hbar^2/(m_\gamma \kbt{})}$ for
ionic species $\gamma$. $\rmupper{\sigma}{in}{\gamma}$ is the intrinsic partition function and will be discussed in detail 
below. From Eq.~\eqref{ionizationbalance} the ionization balance can be described by a Saha equation 
\begin{equation}\label{saha1}
 \frac{n_\alpha \lambda_\alpha^3}{\rmupper{\sigma}{in}{\alpha}}
 = \frac{n_\alpha \lambda_{\alpha+1}^3}{\rmupper{\sigma}{in}{\alpha+1}}\, e^{\beta \rmnumb{\mu}{e}} .
\end{equation}
Since the mass $m_\alpha$ of ion $\alpha$ is almost the same as the mass of its next ionization stage $\alpha+1$, 
i.e. $m_\alpha \approx m_{\alpha +1 }$, we have $\lambda_{\alpha} \approx \lambda_{\alpha+1}$. Therefore,
the ionization equilibrium, Eq.~\eqref{saha1}, can be expressed as
\begin{equation}
 \frac{n_\alpha}{n_{\alpha +1}} 
 = \frac{\rmupper{\sigma}{in}{\alpha}}{\rmupper{\sigma}{in}{\alpha+1}}\, 
  \mathrm{exp} \bks{\beta \rmboth{\mu}{id}{e} + \beta  \rmboth{\Delta}{e}{ionized} }.
\end{equation}

The next step is the determination of the internal partition function $\rmupper{\sigma}{in}{\gamma}$, which
is given by the sum over all possible occupied bound states $\ket{i}$ of the investigated ion species $\gamma$
with $\gamma = \alpha, \alpha+1$
\begin{equation}
 \rmupper{\sigma}{in}{\gamma} = \sum_i^\mathrm{bound}  g_{\gamma i} \, \mathrm{exp} \bks{-\beta\, E_{\gamma i}},
\end{equation}
where $g_{\gamma i} $ is the statistical weight (degeneracy factor). The energy $E_{\gamma i}$ of ion $\gamma$ 
includes the kinetic energy of ion ${\cal E}_{\gamma,\mathbf{p}}$, the internal energy $\varepsilon_{\gamma i}$ 
of a certain configuration $\ket{i}$ of ion describing the internal degrees of freedom in the isolated case, and an 
interaction energy $\Delta_{\gamma i} $ due to correlation with surrounding particles, 
\begin{equation}
 E_{\gamma i} = {\cal E}_{\gamma,\mathbf{p}} + \varepsilon_{\gamma i} + \Delta_{\gamma i} .
\end{equation}
The internal energy of configuration for bound state $\ket{i}$, i.e. $\varepsilon_{\gamma i}$,  
can be rewritten with respect to the ground state of a certain ionic stage $\gamma$. Denoting 
the energy of configuration for the ground state as $\varepsilon_\gamma$, any
other configuration with excitation energy $W_{\gamma i}^0$ with respect to this ground state is then expressed
via $\varepsilon_{\gamma i} = \varepsilon_\gamma + W_{\gamma i}^0$ 
(for the ground state we have $W_{\gamma i}^0=0$). In the chemical picture, the influence 
of charged particle environment on the investigated ion $\gamma$ can be separated into a contribution 
$\Delta_\gamma$ from continuum lowering and a contribution $W_{\gamma i}^1$ accounting for the 
shift of energy level, i.e. $\Delta_{\gamma i} = \Delta_\gamma + W_{\gamma i}^1$. 
Note that we distinguish the terminology of continuum lowering and IPD. The continuum lowering $\Delta_\gamma$
describes the energy change of a structureless particle, whereas IPD contains also structure information
of the ions, such as modification of energy level and coupling between bound electrons and free charged particles.
For energy $E_{\gamma i}$ we have the following relations
\begin{align}
 E_{\gamma i} & = E_\gamma + W_{\gamma i}, \\
 E_\gamma & = {\cal E}_{\gamma,\mathbf{p}} + \varepsilon_{\gamma} + \Delta_{\gamma}, \\
 W_{\gamma i} & = W_{\gamma i}^0 + W_{\gamma i}^1 .
\end{align}
The intrinsic partition function $\rmupper{\sigma}{in}{\gamma} $ can be rewritten as
\begin{equation}
 \rmupper{\sigma}{in}{\gamma}  =  u_\gamma\, e^{-E_\gamma}
\end{equation}
with the standard partition function
\begin{equation}
  u_\gamma = \sum_i^\mathrm{bound}  g_{\gamma i}\, \mathrm{exp} \bks{-\beta\, W_{\gamma i}}.
\end{equation}
Then the Saha equation takes the following form
\begin{equation}
 \frac{n_\alpha}{n_{\alpha +1}}\! =\!\frac{u_\alpha}{u_{\alpha+1}}\,
  \mathrm{exp}\! \squarebra{\beta \rmboth{\mu}{id}{e} \!+\! \beta  \rmboth{\Delta}{e}{ionized}
  \!+ \!\beta \bks{E_{\alpha+1}-E_\alpha} },
\end{equation}
where $E_{\alpha+1}-E_\alpha$ is the energy difference between the ionization stage $\alpha$ and its next
ionization stage $\alpha+1$ and is given by
\begin{align}
 E_{\alpha+1}-E_\alpha = {\cal E}_{\alpha+1,\mathbf{p'}} - {\cal E}_{\alpha,\mathbf{p}}
 + \varepsilon_{\alpha+1} - \varepsilon_{\alpha} + \Delta_{\alpha+1} - \Delta_{\alpha}. \nonumber
\end{align}
Due to the large mass of ions we can assume the momentum of the ion does suffer a slight change, i.e. 
$\mathbf{p} \approx \mathbf{p'}$, so that ${\cal E}_{\alpha+1,\mathbf{p'}} - {\cal E}_{\alpha,\mathbf{p}} \approx 0$.
Obviously, $\varepsilon_{\alpha+1} - \varepsilon_{\alpha}$ is the ionization energy of ionic stage $\alpha$
in the isolated case, which is defined as a positive quantity
\begin{equation}
 I_\alpha = \varepsilon_{\alpha+1} - \varepsilon_{\alpha}.
\end{equation}
We finally obtain the following expression for ionization equilibrium, i.e. the Saha equation,
\begin{equation}
 \frac{n_\alpha}{n_{\alpha +1}} 
 = \frac{u_\alpha}{u_{\alpha+1}}\, \mathrm{exp} \bks{\beta\rmupper{I}{eff}{\alpha} + \beta \rmboth{\mu}{id}{e} }
\end{equation}
with the effective ionization energy 
\begin{equation}\label{ionEneff0}
 \rmupper{I}{eff}{\alpha} = I_\alpha + \bks{ \rmboth{\Delta}{e}{ionized} + \Delta_{\alpha+1}}- \Delta_{\alpha},
\end{equation}
where $\rmboth{\Delta}{e}{ionized} = \rmboth{\Delta}{}{e}+ \rmboth{\Delta}{PF}{bf}$, see Eq.~\eqref{ionEEalpha}.
$\Delta_c$ is the real part of its corresponding SE of particle species $c$.
Evidently, from Eq.~\eqref{ionEneff0} it can be seen that the IPD can be defined 
via the difference between the SE of the corresponding investigated system before and after the ionization.
Such argument supports the definition of IPD within the quantum statistical theory introduced in the Sec.~\ref{DefineIPD}, which is described by
\begin{equation}
\rmboth{\cal I}{IPD}{\alpha} = \Delta_{\alpha} - \bks{ \rmboth{\Delta}{e}{ionized} + \Delta_{\alpha+1}}.
\end{equation}
As already demonstrated in the Sec.~\ref{IPDandSF}, it can be further decomposed into 
\begin{equation}
\rmboth{\cal I}{IPD}{\alpha}   =  \rmupper{\cal I}{sc}{\alpha} + \rmupper{\cal I}{dc}{\alpha} 
\end{equation}
with contribution from the statistical correlations $\rmupper{\cal I}{sc}{\alpha} $ given by 
Eq.~\eqref{ipdSCfinal} and contribution from the dynamical correlations 
$\rmupper{\cal I}{dc}{\alpha} $ described by Eq.~\eqref{ipdINTfinal}.

\section{Results and discussions}
\label{resultsDiss}

\subsection{Weak coupling limit: Debye theory}
\label{DHlimit}

In weakly coupled plasmas, statistical correlation plays a negligible role.
Therefore, the contribution $\rmupper{\cal I}{sc}{\alpha}$ can be ignored
in the calculation. We treat the plasma ions as a whole with effective ionic charge number
$\rmnumb{z}{p}$. Then for the electronic inverse screening length $\rmnumb{\kappa}{e,Debye}$
and total screening parameter $\rmnumb{\kappa}{scr,Debye} = \sqrt{\rmnumb{\kappa}{e,Debye}^2 +\rmnumb{\kappa}{i,Debye}^2}$
we have
\begin{equation}\label{screeningCL}
 \rmnumb{\kappa}{scr,Debye}^2 = \bks{1+ \rmnumb{z}{p}}\, \rmnumb{\kappa}{e,Debye}^2.
\end{equation}
The nonlinear screening function Eq.~\eqref{screening00} for a classical plasma becomes
\begin{equation}
 \rmboth{\kappa}{2}{eff} = 3 \,\Gamma_\alpha 
 = r_\alpha^2\, \rmnumb{\kappa}{scr,Debye}^2.
\end{equation}
For the electronic contribution $\rmboth{\cal I}{dc,el}{\alpha} $, i.e. Eq.~\eqref{eecontrib}, 
we have the following result
\begin{equation}\label{eeContribDebye}
 \rmboth{\cal I}{dc,el}{\alpha} =  \frac{\bks{z_\alpha+1}\, e^2}{4\pi \varepsilon_0 } \, 
 \frac{\rmnumb{\kappa}{scr,Debye}^2}{\rmnumb{\kappa}{e,Debye}}.
\end{equation}
The static SF for ions in a high-temperature ideal plasma is well described by the
Debye expression~\cite{Sperling17}
\begin{equation}
 \rmnumb{S}{ii}(k_0) = \frac{k_0^2 + \rmnumb{\kappa}{e,Debye}^2\, a_0^2}{k_0^2 + \rmnumb{\kappa}{scr,Debye}^2\, a_0^2}
\end{equation}
and the screening cloud is given by
\begin{equation}
 \rmnumb{q}{scr}\bks{k_0} = \frac{ \rmnumb{\kappa}{e,Debye}^2}{k_0^2 + \rmnumb{\kappa}{e,Debye}^2\, a_0^2}.
\end{equation}
The ionic part of interaction contribution in low-density high-temperature plasmas is then evaluated by
\begin{align}\label{iiContribDebye}
 \rmboth{\cal I}{dc,ion}{\alpha}  
 = & \frac{\bks{z_\alpha+1}\, e^2}{2\pi^2 \varepsilon_0 } \,\, a_0\, \rmnumb{\kappa}{scr,Debye}^2 \\
 & \times \int_0^\infty \frac{d k_0}{k_0^2} \, \rmnumb{S}{ii}(k_0)\, \Bigcurlybra{1-\rmnumb{q}{scr}\bks{k_0}}^2 \nonumber \\
 = & \frac{\bks{z_\alpha+1}\, e^2}{4\pi \varepsilon_0 } \, 
 \frac{\rmnumb{\kappa}{scr,Debye}^2}{\rmnumb{\kappa}{scr,Debye} + \rmnumb{\kappa}{e,Debye}}
\end{align}
Inserting the electronic contribution~\eqref{eeContribDebye} and the ionic contribution~\eqref{iiContribDebye} 
into Eq.~\eqref{ipdINTfinal} yields the total interaction contribution of IPD
\begin{equation}
 \rmboth{\cal I}{dc}{\alpha} = \frac{\bks{z_\alpha+1}\, e^2\, \rmnumb{\kappa}{scr,Debye}}{4\pi \varepsilon_0 }
 \cdot L
\end{equation}
with the parameter
\begin{equation}
 L = \frac{1}{1+\rmnumb{z}{p}} \bks{ 
  \frac{\rmnumb{z}{p}\, \rmnumb{\kappa}{scr,Debye}}{\rmnumb{\kappa}{scr,Debye} + \rmnumb{\kappa}{e,Debye}}
  +\frac{\rmnumb{\kappa}{scr,Debye}}{ \rmnumb{\kappa}{e,Debye}}}.
\end{equation}
Using the relation Eq.~\eqref{screeningCL}, it can be shown that $L \equiv 1$. Therefore,
it is demonstrated that the Debye theory for IPD can be perfectly reproduced from our quantum
statistical model based on the SF.

\subsection{Ionization potential depression}
\label{IPDcalculation}
\begin{figure}[h]
\centering
 \includegraphics[width=0.48\textwidth]{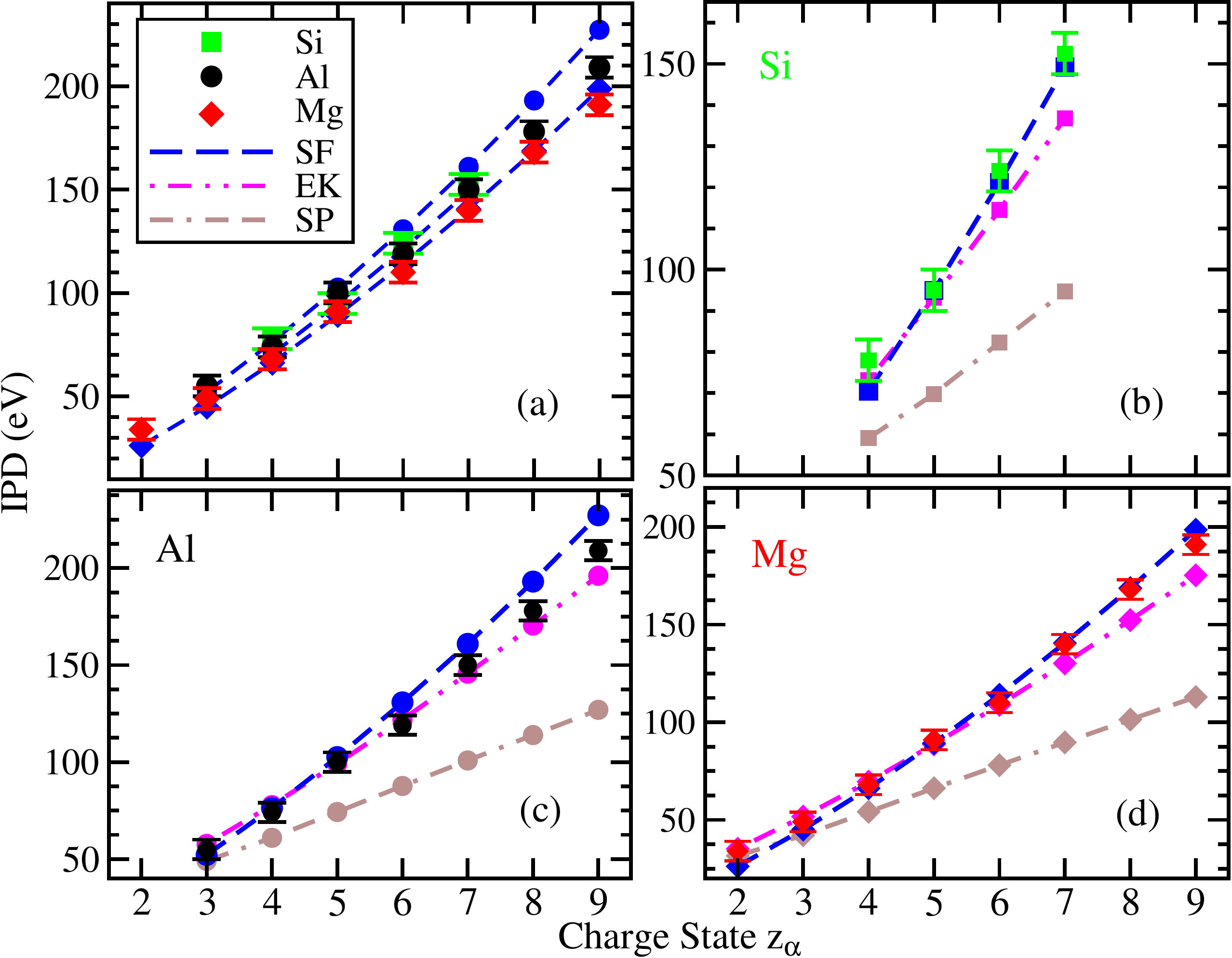}
\caption{IPDs at solid densities for Si (green), Al (black), and Mg (red) as function of charge state $z_\alpha$.  
The experimental values are taken from Ref.~\cite{ciricosta16} with an error bar $\pm\, 5\!$ eV.
 Subfigure (a) shows the experimental results and the predictions of our approaches (SF, blue).
A detailed comparison between the experimental data and predictions from SF (blue), EK (magenta), and SP (brown) model for different elements are shown in subfigure (b) for Si, (c) for Al, and (d) for Mg, respectively.  }
\label{ipdWerte}
\end{figure}

Being a long-standing problem in plasma physics, IPD experiments have been performed recently using the new 
possibility to produce highly excited plasmas near and above condensed matter densities by intense 
short-pulse laser irradiation. In this subsection we discuss the measurements performed by 
Ciricosta \etal~\cite{ciricosta12,ciricosta16}, where the K-edge energies were measured.
By varying the laser photon energy, the trigger energy of the K-shell ionization and the subsequent K$_\alpha$
emission spectra of different charge states in hot dense plasmas were measured and recorded. 
The occurrence of K-shell emission is indeed strongly dependent on the energy of the incident photon, which
can be therefore regarded as an indicator for the direct measurement of the IPDs.
Such measurements were at first performed for aluminum (Al) in 2012 and then in 2016 for other materials such
as magnesium (Mg), silicon (Si) and some chemical compound. The key result extracted from the recorded 
experimental data is that the measured IPDs for Al and Si are insensitive to their environment, even in the 
case of chemical compound. Additionally, for plasmas consisting of electrons and single chemical element,
the IPDs for different charge states have a slight difference for different elements.

\begin{figure}[h]
\centering
 \includegraphics[width=0.48\textwidth]{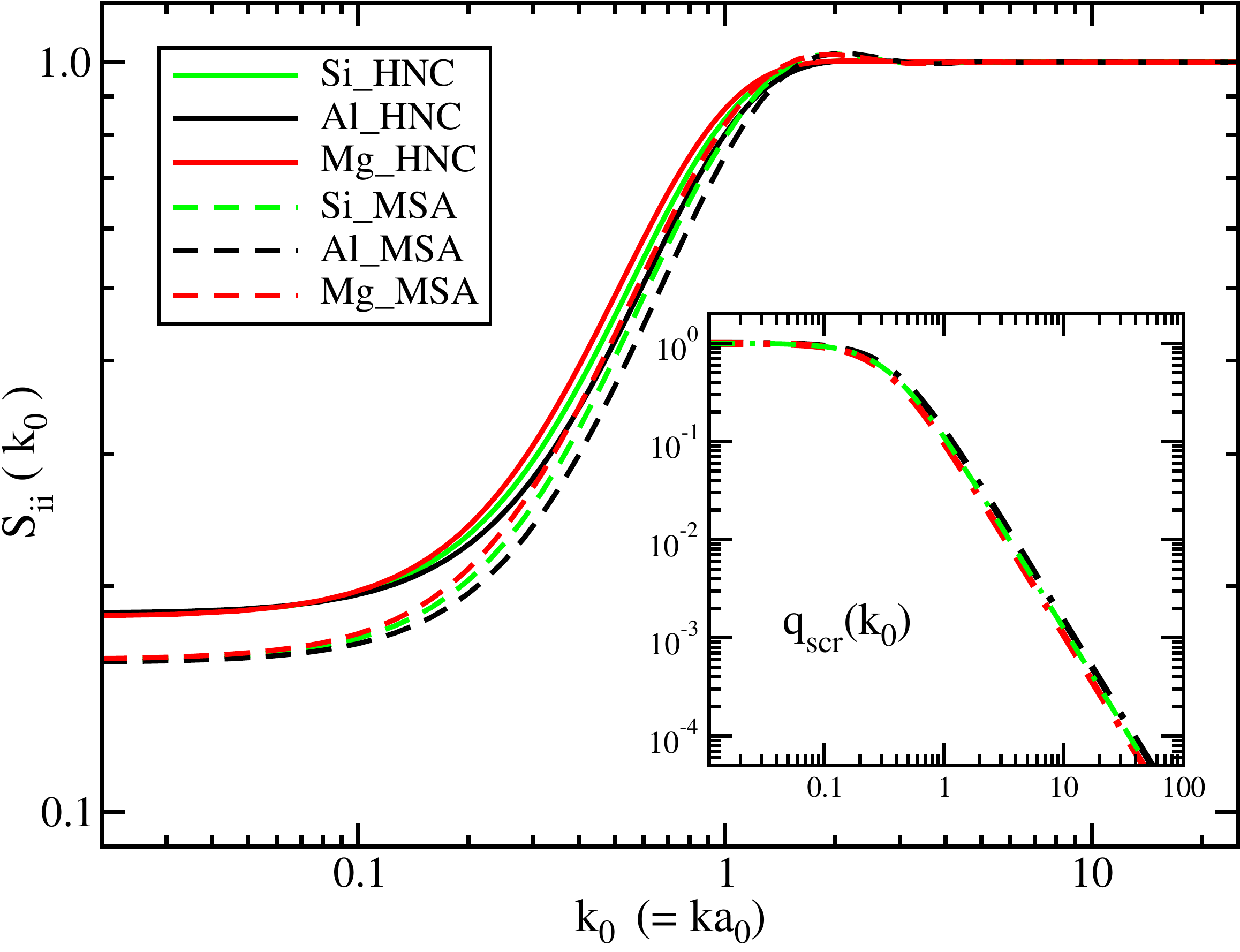}
\caption{Structure factors for Si (green), Al (black), and Mg (red) at solid density with an averaged ionization 
stage $+\, 5$ for electron temperature $\rmnumb{T}{e} = 100 $ eV. For comparison, the HNC method according to 
Ref.~\cite{WHSG08} and fit expression within MSA from Ref.~\cite{GRHGR07} are applied in the calculations of SFs.
The insert describes the corresponding screening function $\rmnumb{q}{scr} (k_0)$, i.e. Eq.~\eqref{screeningLRT}, for 
different elements. }
\label{skfactor}
\end{figure}

Figure~\ref{ipdWerte} shows the experimental results in comparison
to several calculations using different theoretical models.
At solid densities, the corresponding heavy (ion) densities are 
$\rmnumb{n}{\text{\tiny Mg}} = 4.31\times 10^{22}$ cm$^{-3}$,
$\rmnumb{n}{\text{\tiny Al}} = 6.03\times 10^{22}$ cm$^{-3}$, and
$\rmnumb{n}{\text{\tiny Si}} = 4.99\times 10^{22}$ cm$^{-3}$ for Mg, Al, and Si, respectively.
For the calculation we have for the electron temperature $\rmnumb{T}{e} = 100\!$ eV as used in Ref.~\cite{Rosmej18}.
From the comparison with currently experimental data in Fig.~\ref{ipdWerte} (a), it can be seen that 
good agreements for different elements over the whole range of charge states are obtained. 
As already discussed by different authors in Refs.~\cite{ciricosta12,ciricosta16,STJZS14,LRKR17}, 
the SP model fails to explain any of those 
measurements, because the validity of the SP model is restricted to weakly and intermediately coupled plasmas.
The results of EK model match the experimental observations for lower charge states of different elements,
whereas large discrepancies for higher charge states are appeared for all elements~\cite{ciricosta16}.
Such scaling dependence of the charge state $z_\alpha$ can be excellently reproduced
from our quantum statistical approach for elements Mg and Si, as
depicted in Fig.~\ref{ipdWerte} (b) and Fig.~\ref{ipdWerte} (d). For Al plasma as shown in Fig.~\ref{ipdWerte} (c), 
our approach provides slightly larger values $(10 \sim 20 \, \mathrm{eV})$ than the experimental
data, which amounts to $10\%$ of the corresponding experimental results.

In order to have a deep insight on the slight difference of IPDs for different elements, 
as an example, we display the SFs for the plasma ionization with $z_\alpha = \bar z = 5$ in Fig.~\ref{skfactor}.
In our calculation for IPDs we performed the HNC method for the SFs~\cite{WHSG08}. 
For comparison the results of SFs obtained from the fit expression based on the mean spherical 
approximation (MSA)~\cite{GRHGR07} are also shown in the Fig~\ref{skfactor}.
The compressibility of the ion system is described by the ionic SF in the long wavelength limit
$k_0 \rightarrow 0$, which are almost same for different elements at corresponding solid densities
with the values $\rmboth{S}{\text{\tiny HNC}}{ii}(k_0 \rightarrow 0) =  0.1832$
and $\rmboth{S}{\text{\tiny MSA}}{ii}(k_0 \rightarrow 0) =  0.1591$ according to the HNC and MSA 
calculation, respectively. Additionally, for a given wavelength $k_0 < 2$ the MSA always gives a 
little smaller values than the results of HNC. Coming back to the comparison between different elements,
there are a visible difference in the range of $0.1/a_0 \sim 3/a_0$ because of the different ionic densities.
Another factor that affects the IPD values is the screening function $\rmnumb{q}{scr} (k_0)$.
The insert gives the screening function within the linear response framework.
Assuming the same ionization degree, the electron densities for different plasmas at their solid densities are distinct.
Nonetheless, no remarkable discrepancy appears for the screening clouds. 
Therefore, the conclusion drawn from such discussion is twofold. On one hand, similar spatial distribution
of different ionic systems (or effective SF) results in comparable IPD values for these systems. 
On the other hand, the sensitivity of density effects on the IPD can be excellently reflected within our approach
in terms of the SF. Such dependence might be significant for the analysis of the IPD in chemical compounds.

Note that for different charge states of diverse elements the same temperature is utilized in the evaluation
of IPDs. However, the most abundant charge state, and correspondingly the ionization degree (mean charge state),
is generally changed with variation of temperatures.
In the experiments, the temperatures at the time when the average ionization of plasmas equals the charge state
are demonstrated to be different as determined by time-dependent simulations~\cite{ciricosta16}.
Calculations performed with varying temperature for different charge states in the case of Al plasma have been 
reported in our previous work~\cite{LRKR17}. This temperature effect on the IPD and on the ionization 
degree is not intended in this work. Moreover, the ions and the electrons can have different temperatures in owing to 
the short pulse duration in these experiments. We will discuss these effects in association with broadening
of IPD in the forthcoming study.

\subsection{Charge state distribution}
\label{ionicFraction}

To understand the thermodynamic, optical, and transport properties of plasmas, the detailed
knowledge of the charge state distribution is of essential importance. 
As already described in Sec.~\ref{CPandSE}, the charge state distribution can be calculated
by solving the coupled Saha equations incorporating the IPD model. In this section we
consider the ionization balance of aluminum plasmas corresponding to the experiments
performed by Hoarty \etal~\cite{Hoarty13}, where the mass densities and temperatures $(\rho,\rmnumb{T}{e})$ are given as 
following: (1.2 $\pm$ 0.4 g/cc, 550 eV), (2.5 $\pm$ 0.3 g/cc, 650 eV), (5.5 $\pm$ 0.5 g/cc, 550 eV), 
and (9 $\pm$ 1 g/cc, 700 eV). For these measurements
the assumption of local thermodynamic equilibrium is believed to be valid.

\begin{figure}[h]
\centering
 \includegraphics[width=0.48\textwidth]{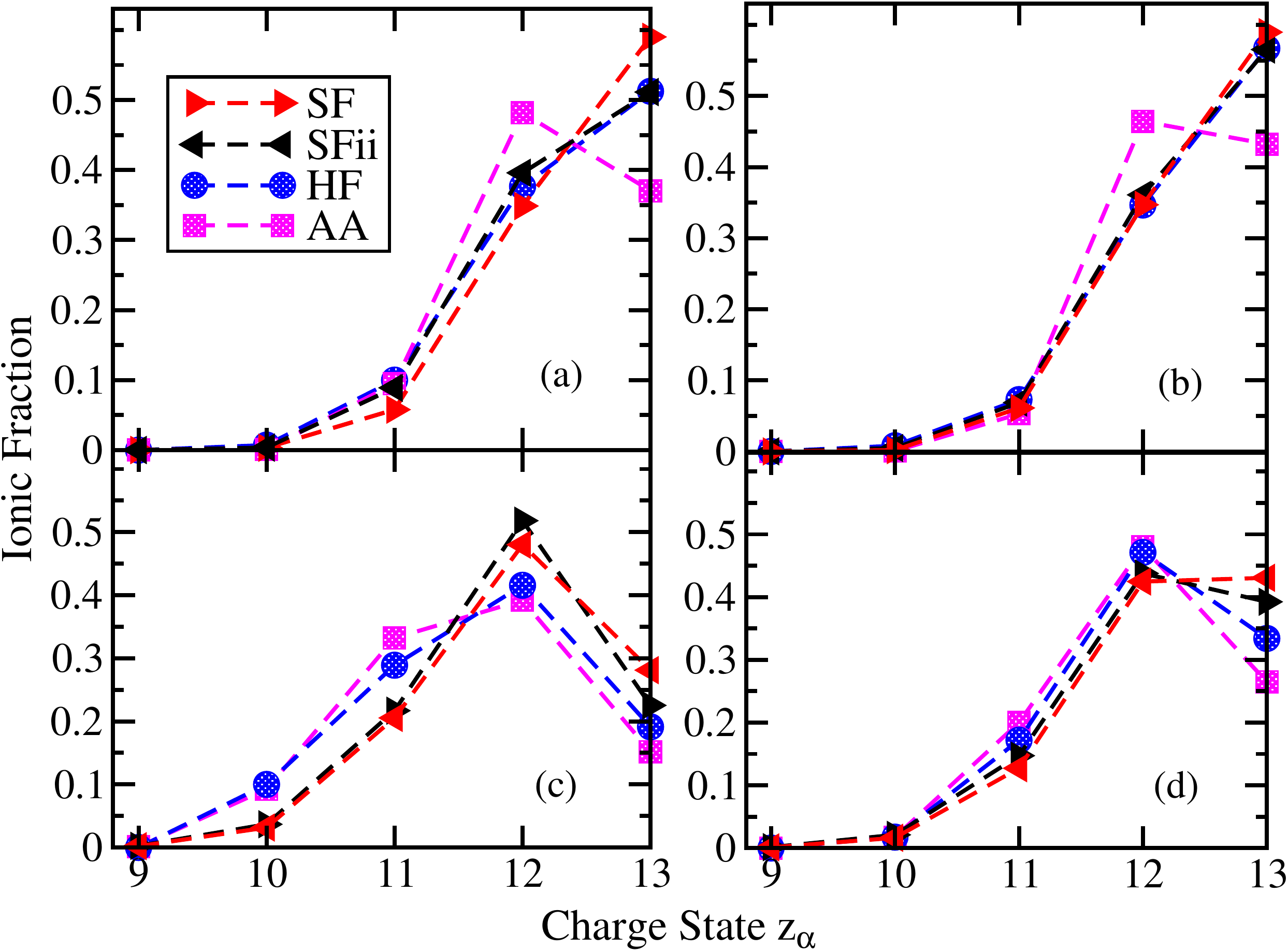}
\caption{Charge state distribution in aluminum plasmas for the following plasma conditions: (a) 1.2 g/cc, 550 eV;
(b) 2.5 g/cc, 650 eV; (c) 5.5 g/cc, 550 eV; (d) 9 g/cc, 700 eV. The results obtained by the average-atom model 
(AA, square symbol) and HF (circle symbol) approaches are taken from Ref.~\cite{FB18}. The predictions denoted by SF 
(triangle right) and SFii (triangle left) are calculated with and without electronic contribution, respectively.}
\label{fraction}
\end{figure}

Fig.~\ref{fraction} highlights the charge state distribution for the above mentioned plasma conditions.
For the low densities of cases (a) and (b), our results show excellent agreements with the 
Hartree-Fock approach~\cite{FB18}, where the mean ionization
is obtained using the configuration occupation probabilities within the framework of Saha-Boltzmann equilibrium.
For higher densities, i.e. cases (c) and (d), the charge state distributions predicted by different approaches 
are quite distinct, although all theoretical models yield the same prediction for the most abundant charge 
state, i.e. ion state Al$^{12+}$. Additionally, according to our theory the electronic
contribution, i.e. Eq.~\eqref{eecontrib},
slightly enhances the ionization degree for all mass density and temperature conditions. For example,
The mean ionizations for such experimental condition are 11.6922, 11.9262, and 12.0064 for the Hartree-Fock approach, 
our SF model without (SFii in Fig.~\ref{fraction}) and with electronic contribution, respectively.

It is remarkable that the prediction of mean ionization depends strongly on the IPD model within
the framework of the Saha-Boltzmann ionization equilibrium. For the experimental condition (c) with
mass density 5.5 g/cc and temperature 550 eV, the large discrepancy among those theoretical models is attributed
to the distinctly predicted IPD values for different charge states.
The binding energies for the level $n=3$ in the ionic stage Al$^{11+} (1s\,3p)$ and Al$^{12+} (3p)$ are
220 eV and 256 eV, respectively. Such levels are pressure ionized in the range of $5.5 \sim 9$ g/cc as 
indicated from the experimental spectra.
As discussed in Ref.~\cite{FB18}, the Hartree-Fock results for synthetic spectra for aluminum  plasma
confirm the predictions of simulations using the SP model of IPD. In the case (c),
the IPD values for Al$^{11+}$ and Al$^{12+}$ are $189$ eV and $200$ eV according to SP model.
The SF model without electronic contribution gives the following IPD values:
$212$ eV for Al$^{11+}$ and $225$ eV for Al$^{12+}$, whereas much larger IPD results are obtained for 
Al$^{11+}$ (294 eV) and Al$^{12+}$ (313 eV) if the electronic contribution are taken into account via the static treatment.
Evidently, the static treatment of free electrons leads to the pressure ionization of $n=3$ levels already in 5.5 g/cc,
which is in contradiction to the experimental observation. Such conflict reveals the importance of dynamical effects of electrons.
Actually, it can be shown that dynamical screening and degeneracy effects come into play already 
in ``weakly'' degenerate plasmas $(\rmnumb{\theta}{ee} \lesssim 10)$, which results in a lowered IPD value in comparison 
to the static treatment of free electrons~\cite{Lin19}.
Furthermore, near the region of pressure ionization fluctuation effects are also of central importance for the determination of
ionization balance.

%
%

\section{Conclusions}
\label{Conclusion}

We have proposed a quantum statistical model for the ionization potential depression in terms of 
the structure factors. Based on the concept of self-energy, a generalized definition for the IPD is introduced,
where not only the shift of continuum edge but also its broadening can be taken into account 
consistently and systematically. Statistical correlations such as quantum exchange and degeneracy effects 
are discussed in detail, whereas the dynamical correlations are reasonably described by the dynamical 
SF by means of the fluctuation-dissipation-theorem. 
The statistical correlations, which are of essential relevance in the high density plasmas,
are generally missing in the commonly applied IPD models.
In particular, the Pauli blocking results in the formation of Fermi surface in highly compressed plasmas~\cite{Hu17},
which strongly modifies the K-edge energy and affects the ionization balance and the optical spectra.
Essentially, the derivation of expression for the IPD does not depend on the assumption of local
thermodynamic equilibrium. Therefore, extension of the developed approach to describe
both LTE and non-LTE plasmas is possible. 

In comparison to the previous work~\cite{LRKR17}, the fit expression for static ionic SF is improved by the HNC
calculations. Furthermore, the proposed IPD model is currently extended to describe multicomponent strongly correlated non-ideal systems in the present work. Additionally, we have also worked out an approach for the calculation of charge state distribution by solving the coupled Saha equations in combination with the developed IPD model.
The validity of our theoretical approach for IPD are also shown, where the Debye-H\"uckel theory
for weakly coupled plasmas can be perfectly reproduced from our method. 
For more strongly coupled plasmas our IPD theory is demonstrated to be suitable to interpret 
the experimental results as shown in the present work and also in our previous study~\cite{LRKR17}.
Density and temperature effects on the IPD are sensitively reflected in the spatial distributions
of particles in plasmas and therefore in the corresponding SFs.
As applications of the developed IPD model, we at first calculated the IPD values at solid
densities for Mg, Al, and Si, where overall good agreements for different elements are shown.
The insensitivity of the measured IPD values for different elements is ascribed to the similarity
of the SFs. Subsequently, the charge state distributions for several density and temperature
conditions are evaluated through the coupled Saha equations, where comparisons with other theoretical
approaches are also performed. Discrepancy in the predication of ionic fractions at the critical density according to different theoretical models reveals that dynamical
screening of free electrons has to be handled carefully~\cite{Lin19}.

A further challenge for the analysis of experimental observations in plasmas is the consideration of broadening effects,
since the discrete eigenstates are broadened to form a band structure in a charged particle system.
Similar to Stark broadening of spectral lines in plasmas, the continuum edge is also broadened due to 
fluctuation effects. Consequently, the IPD is also broadened and can be well described by the imaginary part
of SE in our approach. As a time-averaged effect, the generally discussed IPD do not include the time-dependent
fluctuations. However, the broadening of IPD are sufficiently large to significantly impact the interpretation
of the experimental results~\cite{IS13}. In particular, the broadening effect has to be taken into account cautiously,
in particular in the cases that the IPD values are comparable with the ionization energies. 
Because the energy levels lies near the region of pressure ionization of energy levels, the ionization 
degree is significantly affected by the width of the continuum edge. Therefore, other physical 
properties which depend on the mean ionization $\bar z$ are extremely sensitive to the broadening effect. 
We will extensively discuss the influence of the statistical exchange and broadening effect 
on IPD and on the corresponding ionization balance in a forthcoming work.

\section*{Acknowledgements}
The author gratefully acknowledges much helpful advice from Heidi Reinholz and Gerd R\"opke.
The author also sincerely thanks Yong Hou and Jianmin Yuan for many insightful and fruitful discussions.

\begin{appendix}

\section{Derivation for the dynamical contribution of GIPD}
\label{derivationGIPD}
As shown in the main text,
the dynamical correlation contribution of GIPD $ \rmboth{\cal I}{\,int}{\alpha}(\mathbf{p},\omega) 
=  \rmboth{\cal I}{\,dc}{\alpha}(\mathbf{p},\omega)  + i  \rmboth{\cal B}{\,dc}{\alpha}(\mathbf{p},\omega) $  is given by (see Eqs.~\eqref{realIPD} and~\eqref{imagIPD})
\begin{align}\label{realApp}
 \rmboth{\cal I}{\,dc}{\alpha}(\mathbf{p},\omega)  & =
  {\cal P} \sum_{\mathbf{k}} \int_{-\infty}^{\infty} 
 \frac{d \omega_1}{2\pi} \int_{-\infty}^{\infty}  \frac{d\omega_2}{\pi} \nonumber \\ &\quad \times
 \frac{{\cal A}_\alpha (\mathbf{p,k},\omega_1)}{\omega  - \omega_1 - \omega_2} \cdot 
 \impart\! \! \squarebra{ \frac{ \rmnumb{n}{B}(\omega_2) + 1 }{ \varepsilon(\mathbf{k},\omega_2)} }
\end{align}
and 
\begin{align}\label{imagApp}
 \rmboth{\cal B}{\,dc}{\alpha}(\mathbf{p},\omega)  & =
 - \sum_{\mathbf{k}} \int_{-\infty}^{\infty} \frac{d \omega_1}{2\pi} \,{\cal A}_\alpha (\mathbf{p,k},\omega_1) \nonumber \\ & \quad \times
 \impart\! \! \squarebra{ \frac{ \rmnumb{n}{B}(\omega-\omega_1) + 1 }{ \varepsilon(\mathbf{k},\omega-\omega_1)} } .
\end{align}
The essential quantity in the derivations of Eq.~\eqref{IPDdielectricReal} and Eq.~\eqref{IPDdielectricImag} is
\begin{equation}
Q\bks{\mathbf{p,k},\omega,\omega_2} =  \int_{-\infty}^\infty \frac{d \omega_1}{2\pi} 
 \frac{{\cal A}_\alpha (\mathbf{p,k},\omega_1)}{\omega  - \omega_1 - \omega_2}
\end{equation}
with 
\begin{align}
{\cal A}(&\mathbf{p,k}, \omega_1)
 \! =\! - \frac{2\pi\, \bks{z_\alpha+1}^2 e^2}{\varepsilon_0 k^2}\,
 \delta\!\bks{\hbar \omega_1  \!- \! {\cal E}_{\alpha+1,\mathbf{p-k}}} \\
- &  \frac{2\pi\,  e^2}{\varepsilon_0 k^2}\,\delta\!\bks{\hbar \omega_1 - {\cal E}_{\mathrm{e},\mathbf{p-k}}} 
+ \frac{2\pi\, z_\alpha^2 e^2}{\varepsilon_0 k^2}\,\delta\!\bks{\hbar \omega_1 - {\cal E}_{\alpha,\mathbf{p-k}}} .\nonumber
\end{align}
For the investigated ion $\alpha$, we apply the dispersion relation
$\hbar \omega = {\cal E}_{\alpha,\mathbf{p}}$. Then we arrive at
\begin{align}
& Q\bks{\mathbf{p,k}, {\cal E}_{\alpha,\mathbf{p}}/\hbar,\omega_2}
 = \frac{ z_\alpha^2 e^2}{\varepsilon_0 k^2\, \bks{ {\cal E}_{\alpha,\mathbf{p}}/\hbar -  {\cal E}_{\alpha,\mathbf{p-k}}/\hbar - \omega_2 } } \nonumber \\
& \qquad \qquad \qquad -  \frac{ \bks{z_\alpha+1}^2 e^2}{\varepsilon_0 k^2\,  \bks{ {\cal E}_{\alpha,\mathbf{p}}/\hbar -  {\cal E}_{\alpha+1,\mathbf{p-k}}/\hbar - \omega_2 } } \nonumber \\
&  \qquad \qquad \qquad -  \frac{ e^2}{\varepsilon_0 k^2\,  \bks{ {\cal E}_{\alpha,\mathbf{p}}/\hbar -  {\cal E}_{\mathrm{e},\mathbf{p-k}}/\hbar - \omega_2 } } .
\end{align}
Assuming that the IPD is defined at the momentum $\mathbf{p}=0$~\cite{SAK95}, the propagators in the function 
$Q\bks{\mathbf{p,k}, {\cal E}_{\alpha,\mathbf{p}}/\hbar,\omega_2}$ are reduced to the form
of $1/\squarebra{-\omega'-\hbar k^2/(2m_c)}$.  Taking the classic limit $\hbar \rightarrow 0$ in 
these propagators yields
\begin{align}\label{QQapp}
 Q\bks{\mathbf{0,k}, 0,\omega_2} & =  \frac{ e^2}{\varepsilon_0 k^2\, \omega_2 }
 \squarebra{ 1 + \bks{z_\alpha+1}^2 -z_\alpha^2 } \nonumber \\
 & =   \frac{2 \bks{z_\alpha+1} e^2}{\varepsilon_0 k^2\, \omega_2 }.
\end{align}
Inserting the expression~\eqref{QQapp} into Eq.~\eqref{realApp} and  Eq.~\eqref{imagApp},
we obtain for the shift part of the dynamical correlation contribution
\begin{equation} 
 \rmboth{\cal I}{\, dc}{\alpha} \!\! =\!\! \int \!\!\! \frac{d^3 \mathbf{k}}{\bks{2\pi}^3} \!\!
 \int_{-\infty}^\infty\!\!\! \frac{d \omega_2}{\pi}
 \frac{2\bks{ \! z_\alpha+1 \! } e^2}{\varepsilon_0\, k^2 \, \omega_2} 
 \impart\! \! \squarebra{ \!\frac{ \rmnumb{n}{B}(\omega_2) + 1 }{ \varepsilon(\mathbf{k},\omega_2)}\! },
\end{equation}
and for the broadening contribution
\begin{align}
 \rmboth{\cal B}{\, dc}{\alpha} & =\int \frac{d^3 \mathbf{k}}{\bks{2\pi}^3}
 \int_{-\infty}^\infty d \omega_2 \, \frac{2\bks{z_\alpha+1}\, e^2}{\varepsilon_0\, k^2 } \nonumber \\
 & \quad \times \impart\! \! \squarebra{ \frac{ \rmnumb{n}{B}(\omega_2) + 1 }{ \varepsilon(\mathbf{k},\omega_2)} }\,  \delta\!\bks{\omega_2 + {\cal E}_{\alpha,\mathbf{k}}/\hbar} ,
\end{align}
where the summation $\sum_\mathbf{k}$ is replaced by the integral $\int d^3 \mathbf{k} / \bks{2\pi}^3$.

\section{Charge-charge dynamical SF}
\label{CCDSF}
According to the fluctuation-dissipation theorem,
the effective charge-charge response to an external perturbation, i.e. $\varepsilon({\mathbf k},\omega)$, can be described in terms of the partial density-density dynamical SF $ S_{cd}({\mathbf k},\omega)$
\begin{align}\label{responseApp}
 \impart\! \!  \squarebra{ \frac{1+ \rmnumb{n}{B}(\omega)}{\varepsilon({\mathbf k},\omega) } }
 \!=\!  \frac{\pi e^2}{\hbar \varepsilon_0 k^2} 
 \sum_{cd} z_c z_d \sqrt{n_cn_d} \, S_{cd}({\mathbf k},\omega).
\end{align}
The summation is taken over all particle species in plasmas (i.e. $c,d=$ e, i) and can be rewritten as
\begin{equation}
\sum_{cd} A_{cd} = \rmnumb{A}{ee}+ \sum_\nu \bks{ \rmnumb{A}{\nu e}  + \rmnumb{A}{e \nu} }
+ \sum_{\mu\nu} A_{\mu\nu}.
\end{equation}
Consequently, the expression~\eqref{responseApp} can be reexpressed as
\begin{align}\label{responseApp11}
 \impart\! \!  \squarebra{ \frac{1+ \rmnumb{n}{B}(\omega)}{\varepsilon({\mathbf k},\omega) } }
&= \frac{\pi e^2\, \rmnumb{n}{e}}{\hbar \varepsilon_0 k^2}  
\Bigcurlybra{   \rmnumb{S}{ee}\bks{\mathbf{k},\omega}
\\ & \quad
-   \sum_{\mu} z_\mu \sqrt{\frac{x_\mu }{\bar z}}\,    S_{\mu\mathrm{e}}\bks{\mathbf{k},\omega}
  \nonumber \\ & \quad
 -   \sum_{\nu} z_\nu \sqrt{\frac{x_\nu}{ \bar z}}\,    S_{\mathrm{e}\nu}\bks{\mathbf{k},\omega}
  \nonumber \\ & \quad
+  \sum_{\mu\nu} \frac{z_\mu z_\nu }{\bar z}\sqrt{x_\mu x_\nu} \, S_{\mu\nu}({\mathbf k},\omega)
 \nonumber }.
\end{align}
The electron-ion dynamical SF $ S_{\mathrm{e}\gamma}\bks{\mathbf{k},\omega}$ with
$ S_{\mathrm{e}\gamma}\bks{\mathbf{k},\omega}= S_{\nu \mathrm{e}}\bks{\mathbf{k},\omega}$ and the electron-electron dynamical SF $\rmnumb{S}{ee}\bks{\mathbf{k},\omega}$ are related to 
the ionic dynamical SF $\rmnumb{S}{\gamma\nu}\bks{\mathbf{k},\omega}$ and the free electron
dynamical SF $\rmnumb{S}{ee}^0\bks{\mathbf{k},\omega} $ as follows~\cite{Chihara00,WVGG11}
\begin{align}
 S_{\mathrm{e}\gamma}\bks{\mathbf{k},\omega} 
 & = \sum_\nu \bks{\frac{x_\nu}{\bar{z}}}^{1/2}\,  q_\nu(\mathbf{k}) S_{\gamma\nu}(\mathbf{k}, \omega), \\
\rmnumb{S}{ee}\bks{\mathbf{k},\omega} & = \sum_{\mu\nu} \frac{q_\mu(\mathbf{k})\, q_\nu(\mathbf{k})}{\bar{z}} \bks{x_\mu x_\nu}^{1/2}\,
 S_{\mu\nu}(\mathbf{k}, \omega) \nonumber \\ & \qquad + \rmnumb{S}{ee}^0\bks{\mathbf{k},\omega} .
\end{align}
Inserting these relations into Eq.~\eqref{responseApp11} yields
\begin{align}
& \impart\! \!  \squarebra{ \frac{1+ \rmnumb{n}{B}(\omega)}{\varepsilon({\mathbf k},\omega) } } \\
 =& \frac{\pi e^2\, \rmnumb{n}{e}}{\hbar \varepsilon_0 k^2}   \biggcurlybra{   
\rmnumb{S}{ee}^0\bks{\mathbf{k},\omega}
 +  \sum_{\mu\nu}\frac{z_\mu z_\nu}{\bar z} \sqrt{x_\mu x_\nu} \, S_{\mu\nu}({\mathbf k},\omega)
 \nonumber \\ & \qquad\quad \times 
\squarebra{ 1 - \frac{  q_\mu(\mathbf{k})}{z_\mu } }
\squarebra{ 1 - \frac{  q_\nu(\mathbf{k})}{z_\nu } } \nonumber } \\
= &  \frac{\pi e^2\, \rmnumb{n}{e}}{\hbar \varepsilon_0 k^2}\,  \rmnumb{S}{resp}\bks{\mathbf{k},\omega}
\end{align}
with the total response function
\begin{align}
\rmnumb{S}{resp}\bks{\mathbf{k},\omega} & =  
\rmnumb{S}{ee}^0\bks{\mathbf{k},\omega}
 +  \sum_{\mu\nu}\frac{z_\mu z_\nu}{\bar z} \sqrt{x_\mu x_\nu} \, S_{\mu\nu}({\mathbf k},\omega)
 \nonumber \\ & \qquad\quad \times 
\squarebra{ 1 - \frac{  q_\mu(\mathbf{k})}{z_\mu } }
\squarebra{ 1 - \frac{  q_\nu(\mathbf{k})}{z_\nu } }.
\end{align}
According to the fact that the static SF in the  short-wavelength limit has to be normalized to $1$,
i.e. $S(\mathbf{k}\rightarrow \infty) = 1 $,
so that we introduce a reduced factor in order to ensure the  short-wavelength limit behaviour  of the  total response function $\rmnumb{S}{resp}\bks{\mathbf{k},\omega} $ via the relation
\begin{equation}
\rmnumb{S}{zz}\bks{\mathbf{k},\omega} = \frac{1}{1+\rmnumb{z}{p}} 
\rmnumb{S}{resp}\bks{\mathbf{k},\omega},
\end{equation}
where in the denominator  the effective charge number $\rmnumb{z}{p}$ characterises the charge response of 
the whole ionic mixture, whereas the additional factor $1$ accounts for the charge response of high-frequency
free electrons. Then we obtain the following expression for the fluctuation-dissipation theorem
\begin{align}
 \impart\! \!  \squarebra{ \frac{1+ \rmnumb{n}{B}(\omega)}{\varepsilon({\mathbf k},\omega) } }
 & =  \frac{\pi e^2\, \rmnumb{n}{e} \bks{\rmnumb{z}{p}+1}}{\hbar \varepsilon_0 k^2}\,
 \rmnumb{S}{zz}\bks{\mathbf{k},\omega} \nonumber \\
 & =  \frac{\pi \kbt{}}{\hbar  k^2}\, \rmboth{\kappa}{2}{scr,Debye}\,
 \rmnumb{S}{zz}\bks{\mathbf{k},\omega} .
\end{align}
As discussed in the main text, the Debye screening length is inadequate to describe the
non-linear effects of strongly coupled system.  A modification has to be performed to 
take into account strong coupling effects in plasmas. In this way, the inverse Debye screening parameter
$ \rmboth{\kappa}{}{scr,Debye}$ is replaced by the effective screening parameter
 $ \rmboth{\kappa}{}{scr} =  \rmboth{\kappa}{}{eff}/r_\alpha$ in this work.

\section{Screening theory for impurity-perturber coupling}
\label{screeningTheory}

In a multicomponent plasma the screening cloud around the impurity is given by
\begin{align}
\rmnumb{\rho}{scr} (r)
= \sum_{\nu} (z_\nu e)\, n_{z  \nu}(r) - e\, n_{z\mathrm{ e}}(r) .
\end{align}
with the charge distribution
\begin{equation}
 n_{zj}(r) 
= n_j\, \mathrm{exp } \bks{ - \frac{z_j e\, \psi_z(r)}{\kbt{}} },
\end{equation}
where the electrostatic potential reads
\begin{equation}\label{potenApp}
 \psi_z(r) = \frac{z e}{4\pi\varepsilon_0 \, r} \, \mathrm{exp }\bks{- \rmnumb{\kappa}{scr} r}.
\end{equation}
To determine the screening cloud and the corresponding screening parameter, detailed knowledge of the 
charge state distribution is necessary and integration over a series of transcendental functions has to be
performed. To simplify the calculation, we can use the concept of effective perturber with charge number $\rmnumb{z}{p}$, 
which effectively describes the property of the plasma as a whole. The screening cloud $\rmnumb{\rho}{scr} (r)$ can 
be approximated as follows
\begin{align}
&\rmnumb{\rho}{scr} (r)
 = \sum_{\nu} (z_\nu e)\, n_{z  \nu}(r) - e\, n_{z\mathrm{ e}}(r) \nonumber \\
& \approx  e \sum_{\nu} z_\nu\,n_\nu\, \bks{ 1 -  \frac{z_\nu e\, \psi_z(r)}{\kbt{}} }  - e \rmnumb{n}{e} \bks{ 1 + \frac{e\, \psi_z(r)}{\kbt{}} }.
\end{align}
Due to the charge neutrality $ \sum_{\nu} z_\nu n_\nu - \rmnumb{n}{e} = 0$, we have
\begin{equation}
\rmnumb{\rho}{scr} (r) =  - 
\frac{e^2\,  \psi_z(r)}{\kbt{}}\biggcurlybra{ \sum_{\nu} z_\nu^2 n_\nu + \rmnumb{n}{e}\! }.
\end{equation}
Using the definition $\rmnumb{z}{p}= \sum_{\nu} z_\nu^2 n_\nu/ \rmnumb{n}{e}$, the following expression can be obtained
for the screening cloud
\begin{align}
\rmnumb{\rho}{scr} (r) & = - \frac{e^2\, \rmnumb{n}{e}\, \bks{ \rmnumb{z}{p} + 1 } \,\psi_z(r)}{\kbt{}} \\
& = e \rmnumb{n}{e} \bks{ 1 - \frac{ \bks{\rmnumb{z}{p}+1} e  \psi_z(r) }{\kbt{}} -1 } \nonumber \\
& \approx  e \rmnumb{n}{e} \bks{ \mathrm{exp}\!\bks{ - \frac{ \bks{\rmnumb{z}{p}+1} e  \psi_z(r) }{\kbt{}} } -1 }.
\end{align}
%
This expression describes the screening for the impurity-perturber coupling if we treat the plasma as a whole.
Considering the ionization reaction and the relaxation of charge distribution, we take the impurity as 
the ion after ionization, i.e. $z = z_\alpha + 1$. Inserting Eq.~\eqref{potenApp} into Eq. yields
\begin{equation}
 \rmnumb{\rho}{scr}  (r)  \!
 = \! e \rmnumb{n}{e} \! \curlybra{\!\! \mathrm{exp}\!\!
 \squarebra{\! - \frac{\bks{z_\alpha \!+ \! 1}  \bks{\rmnumb{z}{p} \!+ \! 1} e^2}
 {4\pi \varepsilon_0 r \kbt{} }\, e^{-r\rmnumb{\kappa}{scr}} \! } \!\!\!-\!1\! }.
\end{equation}
Using the condition of charge neutrality for the screening cloud $z e + \int d^3 \mathbf{r} \rmnumb{\rho}{scr}  (r)  =0$,
the screening parameter $\rmnumb{\kappa}{scr}$ can be determined.

\end{appendix}

\end{document}